\documentclass[aps,preprint,amsmath,amssymb,amsfonts]{revtex4}
\usepackage{epsfig}
\usepackage{graphicx}
\usepackage{subfigure}
\usepackage{dcolumn}
\usepackage{bm}
\usepackage{amsthm}
\usepackage{enumitem}
\usepackage{slashed}
\usepackage{braket}
\usepackage{amsmath}

\newcommand{\del}{\partial}

\renewcommand{\Re}{\operatorname{Re}}

\newcommand{\tr}{\operatorname{tr}}

\newcommand{\bbR}{\mathbb{R}}

\newcommand{\calB}{\mathcal{B}}

\newcommand{\calE}{\mathcal{E}}

\newcommand{\calK}{\mathcal{K}}

\newcommand{\calN}{\mathcal{N}}
\newcommand{\calO}{\mathcal{O}}

\begin{document}

\title{Bulk interactions and boundary dual of higher-spin-charged particles}

\author{Adrian David}
\email{adrian.david@oist.jp}
\affiliation{Okinawa Institute of Science and Technology, 1919-1 Tancha, Onna-son, Okinawa 904-0495, Japan}
\author{Yasha Neiman}
\email{yashula@icloud.com}
\affiliation{Okinawa Institute of Science and Technology, 1919-1 Tancha, Onna-son, Okinawa 904-0495, Japan}

\date{\today}

\begin{abstract}
We consider higher-spin gravity in (Euclidean) $AdS_4$, dual to a free vector model on the 3d boundary. In the bulk theory, we study the linearized version of the Didenko-Vasiliev black hole solution: a particle that couples to the gauge fields of all spins through a BPS-like pattern of charges. We study the interaction between two such particles at leading order. The sum over spins cancels the UV divergences that occur when the two particles are brought close together, for (almost) any value of the relative velocity. This is a higher-spin enhancement of supergravity's famous feature, the cancellation of the electric and gravitational forces between two BPS particles at rest. In the holographic context, we point out that these ``Didenko-Vasiliev particles'' are just the bulk duals of bilocal operators in the boundary theory. For this identification, we use the Penrose transform between bulk fields and twistor functions, together with its holographic dual that relates twistor functions to boundary sources. In the resulting picture, the interaction between two Didenko-Vasiliev particles is just a geodesic Witten diagram that calculates the correlator of two boundary bilocals. We speculate on implications for a possible reformulation of the bulk theory, and for its non-locality issues. 
\end{abstract}

\maketitle
\tableofcontents
\newpage

\section{Introduction and summary} \label{sec:intro}

Higher-spin (HS) gravity \cite{Vasiliev:1995dn,Vasiliev:1999ba} is a smaller sibling of string theory. It is a theory of infinitely many massless fields with different spins, including a spin-2 ``graviton''. Whereas string theory is holographically dual to matrix-like conformal field theories \cite{Maldacena:1997re,Witten:1998qj,Aharony:1999ti}, higher-spin gravity is dual to vector models \cite{Klebanov:2002ja,Sezgin:2002rt,Sezgin:2003pt,Giombi:2012ms}. While HS gravity can be defined in various dimensions, our interest will be the 4d case. Specifically, we will consider the type-A theory, which has one field of every integer spin, and is holographically dual to a free vector model of $N$ complex scalar fields \cite{Klebanov:2002ja}.

From a complementary point of view, HS gravity is a \emph{larger} sibling of supergravity, the low-energy limit of string theory. Whereas supergravity extends the spacetime symmetry of GR in a fermionic direction, HS gravity extends it in a bosonic direction, resulting in an infinite multiplet of massless ``partners'' for the graviton with different spins. In supergravity with $\calN=2$ or higher supersymmetry, particular importance is placed on extremal black hole (or black brane) solutions \cite{Horowitz:1991cd} that saturate the BPS bound. Such black holes define backgrounds that preserve part of the theory's supersymmetry. The gravitational charge of these black holes, i.e. their mass, is proportional to their electric charge under gravity's spin-1 superpartner. These supergravity solutions play a key role in string theory, where they are ultimately identified \cite{Polchinski:1995mt} with D-branes \cite{Dai:1989ua,Horava:1989ga}.

In HS gravity, a similar object is known -- the Didenko-Vasiliev ``black hole'' \cite{Didenko:2009td}. This is a spherically symmetric solution to the Vasiliev equations, constructed by analogy with the Kerr-Schild procedure for the Schwarzschild black hole. Aside from this formal similarity, it's not at all clear that this solution behaves like the familiar black holes of GR, with their event horizons and thermodynamical properties; hence the quotation marks around the term ``black hole''.  See  \cite{Iazeolla:2011cb} for generalizations of the Didenko-Vasiliev solution, as well as \cite{Iazeolla:2017vng}, where general HS-extended Petrov Type D solutions are interpreted as black hole microstates. The Didenko-Vasiliev black hole is charged under the HS fields of all spins. These charges are all proportional to each other, reflecting a partial preservation of HS symmetry, in clear analogy with the supergravity case. Though we won't consider it here, there is also a supersymmetric version of the Didenko-Vasiliev solution, where the type-A and type-B HS gravities are naturally combined into an $\calN=2$ supermultiplet. One then finds \cite{Didenko:2009td} that the solution preserves a quarter of the supersymmetries, further strengthening the analogy to BPS black holes. 

In this paper, we will study the linearized version of the Didenko-Vasiliev black hole, which was described by the same authors in \cite{Didenko:2008va}. This is a solution to Fronsdal's equations for free HS gauge fields \cite{Fronsdal:1978rb,Fronsdal:1978vb}, with a source localized at the spatial origin (or, from a spacetime perspective, along the time axis). We will refer to this source as a ``Didenko-Vasiliev particle'', or DV particle for short. Our main case of interest, and the original context of \cite{Didenko:2008va}, will be DV particles in $AdS_4$, where a non-linear HS theory exists, complete with holographic duals. However, we will also consider DV particles in flat spacetime, where some calculations simplify. We will work in Euclidean signature, so that our spacetime will be either flat $\bbR^4$ or hyperbolic space, i.e. Euclidean $AdS_4$ ($EAdS_4$ for short). 

The main object of our calculations will be the leading-order action from two DV particles interacting via their HS gauge fields. For every spin separately, this action is IR-divergent in $\bbR^4$, but finite in $EAdS_4$. When the particles are brought close together, a UV divergence arises; this is the same as the IR divergence from the flat case, viewed from a different perspective. However, we find that these divergences cancel upon summing over spins, thanks to the DV particles' special pattern of charges. This is an enhanced version of the well-known cancellation of the electric and gravitational forces between two BPS objects in supergravity (or between two extremal Riessner-Nordstrom black holes). In the latter case, the cancellation only holds when the two objects are mutually at rest; the introduction of a relative velocity reveals the different tensor structure of the two forces, and they no longer cancel. In our case, the cancelation holds for any relative velocity, with the single exception of a particle and antiparticle mutually at rest. This criterion can in fact be used to \emph{derive} the DV pattern of charges (we are grateful to Slava Lysov for calling our attention to this feature). Either way, it reflects a certain kind of non-locality, or softness, of interactions governed by HS symmetry.

Our next observation is that the higher-spin fields of the DV particle, as given in \cite{Didenko:2008va}, play a role in higher-spin holography. Since the original motivation in \cite{Didenko:2008va,Didenko:2009td} was to mimic black hole solutions of GR, one might expect that the relevant holographic duality would be between non-perturbative black holes in the bulk and thermal states in the boundary CFT. We make no claims here about the validity of this scenario. Instead, we point out a different one, in which the \emph{linearized} DV solution appears in a \emph{perturbative} role. Specifically, the DV particle describes the linearized bulk solution, i.e. the ``boundary-to-bulk propagator'', that corresponds to a bilocal operator \cite{Das:2003vw,Douglas:2010rc} in the boundary theory. In this picture, the worldline of the DV particle, which sources its HS fields, lies on the bulk geodesic that connects the two boundary ``legs'' of the bilocal -- see figure \ref{fig:3pt}. Note that this can't directly apply in the original setup of \cite{Didenko:2008va,Didenko:2009td}: there, the DV particle's worldline is a timelike geodesic in $AdS_4$, which has no boundary endpoints. The problem is not fatal, though: a timelike $AdS_4$ geodesic \emph{does} have boundary endpoints in \emph{complexified} spacetime, so we can consider a bilocal boundary operator there. Alternatively, we can work in de Sitter space, where timelike geodesics have real boundary endpoints. In this paper, we'll be working in $EAdS_4$, where all geodesics are spacelike, and have real boundary endpoints. For a systematic discussion of various signatures in higher-spin theory, see \cite{Iazeolla:2007wt}.

It is instructive to compare our picture to the one developed in e.g. \cite{Iazeolla:2017vng,Iazeolla:2020jee}. There, a distinction is made between ``particle-like'' and ``black-hole-like'' HS field solutions. In this terminology, what we refer to as a DV particle falls into the ``black-hole-like'' category. In particular, one shouldn't confuse the DV particle, a spin-0 particle-like object charged under the HS fields, with a particle excitation of the HS fields themselves! On the other hand, in our picture, there is a continuous limiting procedure that relates HS field solutions \emph{with} a DV-like source to ones \emph{without}: one must simply bring together the DV worldline's boundary endpoints. This will reduce the bilocal boundary operator to the standard HS tower of local currents, and the DV bulk solution to an ordinary boundary-to-bulk propagator. Thus, the linearized HS fields with DV sources can be regarded as a \emph{continuous generalization} of the source-free solutions -- a point of view that we'll adopt in section \ref{sec:discuss}. This should be contrasted with the picture in \cite{Iazeolla:2017vng}, where the ``particle-like'' and ``black-hole-like'' solutions are associated with distinct representations of the spacetime symmetry, and cannot mix with each other. The apparent contradiction is resolved by being careful about the relevant spacetime signatures. The distinction between singleton and anti-singleton representations, which plays a defining role in \cite{Iazeolla:2017vng}, is only applicable in Lorentzian $AdS_4$, with its $SO(2,3)$ symmetry. As mentioned above, the relevant boundary endpoints in that case are complex, and they cannot be brought together without changing the causal character of the DV particle's worldline. In our Euclidean $AdS_4$ setup, the boundary endpoints are real, and can be brought together without any problem.

Our identification of the DV particle with a boundary bilocal operator is obtained via the spacetime-independent twistor formalism for HS holography, developed in \cite{Neiman:2017mel,David:2020ptn}. Though the appearance of a particle-like bulk source in the context of a ``boundary-to-bulk propagator'' may sound surprising, it is in fact a special case of the recently uncovered relation \cite{Hijano:2015zsa,daCunha:2016crm,Chen:2019fvi} between OPE blocks in the boundary CFT and geodesic Witten diagrams. In particular, the interaction of two bulk DV particles in $EAdS_4$, which we will calculate here, is nothing but a geodesic Witten diagram for the correlator of two boundary bilocals. 

The rest of the paper is structured as follows. In section \ref{sec:Fronsdal}, we review Fronsdal's theory of free HS fields, in flat space and $EAdS_4$, including boundary-to-bulk propagators and 2-point functions. In section \ref{sec:bulk_particle}, we construct and solve the field equations for linearized HS fields sourced by a bulk particle that travels along a geodesic worldline. In section \ref{sec:two_particles}, we study the action for two such particles interacting via HS fields, and demonstrate the cancellation of UV divergences for the DV pattern of charges. In the flat case, we find the action analytically; in $EAdS_4$, we end up with an unpleasant integral, which however agrees numerically with a simple analytic answer guessed from holography.

In section \ref{sec:twistors}, we take a detour to introduce $EAdS_4$ twistors, higher-spin algebra, the Penrose transform, as well as HS-algebraic formulas for boundary-to-bulk propagators and boundary $n$-point functions. In this, we will follow the formalism of \cite{Neiman:2013hca,Neiman:2017mel,David:2020ptn}, which combines HS algebra, embedding space, and spacetime-independent twistors. With this machinery in place, we proceed to make our main claims in section \ref{sec:boundary_bilocals}. There, we evaluate the linearized bulk fields that correspond to a bilocal operator in the boundary CFT, and notice that they coincide with the fields of a DV particle. We further notice that the interaction of two such particles, as calculated in section \ref{sec:two_particles}, reproduces the correlator of two boundary bilocals, in the spirit of the general theory of geodesic Witten diagrams \cite{Hijano:2015zsa}. Finally, in section \ref{sec:discuss}, we speculate about the relevance of our construction to understanding interacting HS theory and its locality issues.

\section{Free HS fields and local boundary sources} \label{sec:Fronsdal}

In this section, we review the theory of free HS gauge fields, before introducing HS-charged particles in section \ref{sec:bulk_particle}.

\subsection{Fronsdal action, field equations and gauge symmetry} \label{sec:Fronsdal:equations}

The theory of linearized higher-spin gauge fields on maximally symmetric spacetimes, with or without cosmological constant, was put forward by Fronsdal in \cite{Fronsdal:1978rb,Fronsdal:1978vb}. It generalizes Maxwell theory (which constitutes the spin-1 case) and linearized GR (the spin-2 case). Also included is the spin-0 case of a conformally massless scalar, though it is not strictly speaking a gauge theory.

We will work in 4d Euclidean spacetime with a positive-definite metric $g_{\mu\nu}$. The specific spacetime will be either flat $\bbR^4$, or Euclidean $AdS_4$ of unit radius. The commutator of covariant derivatives $\nabla_\mu$ in these spacetimes reads:
\begin{align}
  [\nabla_\mu,\nabla_\nu]v^\rho = \left\{\begin{array}{cc}
    0 & \quad \bbR^4 \\  
   2v_{[\mu}\delta_{\nu]}^\rho & \quad EAdS_4
  \end{array} \right. \ .
\end{align} 
A spin-$s$ gauge potential in Fronsdal's formulation is given by a tensor $h_{\mu_1\dots\mu_s}$ that is totally symmetric and double-traceless in its indices:
\begin{align}
 h_{\mu_1\dots\mu_s} = h_{(\mu_1\dots\mu_s)} \ ; \quad h^{\nu\rho}_{\nu\rho\mu_1\dots\mu_{s-4}} = 0 \ .
\end{align} 
The first of these constraints becomes non-trivial for $s\geq 2$, and the second -- for $s\geq 4$. The low-spin cases $h$, $h_\mu$ and $h_{\mu\nu}$ correspond respectively to a scalar field, a Maxwell potential and a linearized metric perturbation. For $s\geq 1$, the fields $h_{\mu_1\dots\mu_s}$ are subject to a gauge symmetry:
\begin{align}
  h_{\mu_1\dots\mu_s} \ \rightarrow \ h_{\mu_1\dots\mu_s} + \nabla_{(\mu_1}\theta_{\mu_2\dots\mu_s)} \ , \label{eq:gauge_transformation}
\end{align}
where the gauge parameter $\theta_{\mu_1\dots\mu_{s-1}}$ is in turn constrained to be totally symmetric and traceless (a constraint that becomes non-trivial for $s\geq 3$):
\begin{align}
 \theta_{\mu_1\dots\mu_{s-1}} = \theta_{(\mu_1\dots\mu_{s-1})} \ ; \quad \theta^\nu_{\nu\mu_1\dots\mu_{s-3}} = 0 \ . \label{eq:theta_constraint}
\end{align}
One can construct from $h_{\mu_1\dots\mu_s}$ a gauge-invariant second-derivative object, known as the Fronsdal tensor. In flat space, this reads:
\begin{align}
 F_{\mu_1\dots\mu_s} = \Box h_{\mu_1\dots\mu_s} - s\nabla_{(\mu_1}\nabla^\nu h_{\mu_2\dots\mu_s)\nu} + \frac{s(s-1)}{2}\nabla_{(\mu_1}\nabla_{\mu_2}h^\nu_{\mu_3\dots\mu_s)\nu} \ , \label{eq:F_raw_flat}
\end{align}
where $\Box = \nabla_\mu\nabla^\mu$. In $EAdS_4$, we get additional terms due to the curvature:
\begin{align}
 \begin{split}
   F_{\mu_1\dots\mu_s} ={}& (\Box + 2 - 2s^2)h_{\mu_1\dots\mu_s} - s\nabla^\nu\nabla_{(\mu_1}h_{\mu_2\dots\mu_s)\nu} + \frac{s(s-1)}{2}\nabla_{(\mu_1}\nabla_{\mu_2}h^\nu_{\mu_3\dots\mu_s)\nu} \\
     ={}& (\Box + 2 + 2s - s^2)h_{\mu_1\dots\mu_s} - s\nabla_{(\mu_1}\nabla^\nu h_{\mu_2\dots\mu_s)\nu} + \frac{s(s-1)}{2}\nabla_{(\mu_1}\nabla_{\mu_2}h^\nu_{\mu_3\dots\mu_s)\nu} \\
       &- s(s-1)g_{(\mu_1\mu_2}h^\nu_{\mu_3\dots\mu_s)\nu} \ ,
 \end{split} \label{eq:F_raw_EAdS}
\end{align}
where the difference between the two expressions is in the ordering of the derivatives in the second term. For $s=1$, $F_\mu$ is the divergence of the Maxwell field strength; for $s=2$, $F_{\mu\nu}$ is proportional to the linearized Ricci tensor. The free field equations for all spins are simply $F_{\mu_1\dots\mu_s} = 0$. 

The linear equations of motion $F_{\mu_1\dots\mu_s} = 0$ can be derived from a quadratic Lagrangian of the form $\sim h_{\mu_1\dots\mu_s}F^{\mu_1\dots\mu_s}$. However, for $s\geq 2$, this Lagrangian is not invariant under the gauge transformation \eqref{eq:gauge_transformation}. We must instead use a trace-modified version of $F_{\mu_1\dots\mu_s}$:
\begin{align}
 G_{\mu_1\dots\mu_s} = F_{\mu_1\dots\mu_s} - \frac{s(s-1)}{4}\,g_{(\mu_1\mu_2} F^\nu_{\mu_3\dots\mu_s)\nu} \ ; \quad F_{\mu_1\dots\mu_s} = G_{\mu_1\dots\mu_s} - \frac{s}{4}\,g_{(\mu_1\mu_2} G^\nu_{\mu_3\dots\mu_s)\nu} \ , \label{eq:F_G}
\end{align}
which satisfies a (partial) conservation law:
\begin{align}
 \nabla_\nu G^\nu_{\mu_1\dots\mu_{s-1}} = \text{trace terms only} \ . \label{eq:G_conservation}
\end{align}
For $s=2$, this is just the construction of the Einstein tensor out of the Ricci tensor. For spins 1 and 2, $G_{\mu_1\dots\mu_s}$ is conserved as usual. For $s\geq 3$, it is not the full divergence $\nabla_\nu G^\nu_{\mu_1\dots\mu_{s-1}}$ that vanishes, but rather its \emph{totally traceless part}. This is sufficient to define an action:
\begin{align}
 S = -\frac{1}{2}\int d^4x\sqrt{g}\,h_{\mu_1\dots\mu_s}G^{\mu_1\dots\mu_s} + \text{boundary terms} \ , \label{eq:S}
\end{align}
which is invariant (up to boundary terms) under the gauge transformation \eqref{eq:gauge_transformation} subject to the constraint \eqref{eq:theta_constraint} on the gauge parameter. The Euler-Lagrange equations of motion are now $G_{\mu_1\dots\mu_s} = 0$, which is of course equivalent to $F_{\mu_1\dots\mu_s} = 0$. For more detailed discussion of free HS Lagrangians, see e.g. \cite{Curtright:1979uz,Buchbinder:2001bs}.

For solutions to the source-free field equations $F_{\mu_1\dots\mu_s} = 0$, it's possible to choose a transverse traceless gauge. The gauge conditions and field equations can then be summarized as:
\begin{align}
 h^\nu_{\nu\mu_1\dots\mu_{s-2}} = 0 \ ; \quad \nabla^\nu h_{\nu\mu_1\dots\mu_{s-1}} = 0 \ ; \quad \Box h_{\mu_1\dots\mu_s} = m^2 h_{\mu_1\dots\mu_s} \ , \label{eq:TT_eqs}
\end{align}
where $m^2 = 0$ in flat space, and $m^2 = s^2 - 2s - 2$ in $EAdS_4$. 

The gauge-invariant content of a \emph{solution} to the field equations is captured by another invariant tensor, this time involving $s$ derivatives:
\begin{align}
 \begin{split}
  \varphi_{\mu_1\nu_1\dots\mu_s\nu_s} = 2^s\,\nabla_{\mu_1}\dots\nabla_{\mu_s} h_{\nu_1\dots\nu_s} \quad &\text{antisymmetrized over every }\mu_k\nu_k\text{ pair,} \\
   &\text{with all traces subtracted.}
 \end{split} \label{eq:C}
\end{align}
For $s=0$, this is just the scalar field again, $\varphi=h$; for $s=1$, $\varphi_{\mu\nu}$ is the Maxwell field strength; for $s=2$, $\varphi_{\mu\nu\rho\sigma}$ is proportional to the linearized Weyl tensor. The field strength \eqref{eq:C} is the sum of two chiral parts: one that is right-handed (i.e. self-dual) in every $\mu_k\nu_k$ index pair, and one that is left-handed (i.e. anti-self-dual). From the point of view of the field strength \eqref{eq:C}, without referring to the gauge potential $h_{\mu_1\dots\mu_s}$, the free massless field equations read:
\begin{align}
 \begin{split}
   s = 0:\quad &\Box\varphi = m^2\varphi \ ; \\
   s = 1:\quad &\nabla^\mu \varphi_{\mu\nu} = \nabla_{[\mu} \varphi_{\nu\rho]} = 0 \ ; \\
   s\geq 2: \quad &\nabla^{\mu_1} \varphi_{\mu_1\nu_1\cdots\mu_s\nu_s} = 0 \ ,
 \end{split} \label{eq:C_field_eqs}
\end{align}
where the only difference between $\bbR^4$ and $EAdS_4$ is now in the spin-0 case, with $m^2=0,-2$ respectively.

\subsection{Boundary data and on-shell action in $EAdS_4$} \label{sec:Fronsdal:boundary}

We now specialize to $EAdS_4$ spacetime. In this subsection, we will represent $EAdS_4$ in Poincare coordinates $(z,\mathbf{x})$, while writing the components of tensors in an orthonormal basis. This is described by the vielbein:
\begin{align}
 e_z^0 = e_1^1 = e_2^2 = e_3^3 = \frac{1}{z} \ .
\end{align}
All tensor indices will refer to the orthonormal basis. Greek indices $(\mu,\nu,\dots)$ will take values in $(0,1,2,3)$, while Latin indices $(i,j,\dots)$ take values in $(1,2,3)$. 

The conformal boundary of $EAdS_4$, here in a flat conformal frame, is at $z=0$. The boundary behavior of solutions to the Fronsdal equations has beed studied e.g. in \cite{Mikhailov:2002bp,Joung:2011xb,Campoleoni:2016uwr}. The results are simplest in transverse traceless gauge, where the equations take the form \eqref{eq:TT_eqs}. As usual, locally near the boundary, there are two branches of linearly independent solutions, characterized by different powers of $z$, which are canonically conjugate to each other. Global regularity on $EAdS_4$ picks out a particular linear combination of the two branches. We will refer to the branches as ``electric'' and ``magnetic'', for reasons that will become clear. Their asymptotics, at leading order in small $z$, is defined by:
\begin{align}
  \text{Magnetic branch}: \quad &h_{i_1\dots i_s}(z,\mathbf{x}) = z^{2-s} A_{i_1\dots i_s}(\mathbf{x}) \ ; \label{eq:magnetic_branch} \\
  \text{Electric branch}:\quad &h_{i_1\dots i_s}(z,\mathbf{x}) = z^{s+1} J_{i_1\dots i_s}(\mathbf{x}) \ . \label{eq:electric_branch}
\end{align}
The other components of $h_{\mu_1\dots\mu_s}$ within each of the branches, i.e. the components where one or more indices take the value 0, scale with higher powers of $z$, and are determined by the components $h_{i_1\dots i_s}$ above. The tracelessness of $h_{\mu_1\dots\mu_s}$ then implies that $A_{i_1\dots i_s}$ and $J_{i_1\dots i_s}$ must be traceless. In addition, the electric boundary data $J_{i_1\dots i_s}$ is divergence-free, $\del_{i_1}J^{i_1 i_2\dots i_s} = 0$. In the holographic duality with a free vector model of scalar fields, the electric boundary data $J_{i_1\dots i_s}(\mathbf{x})$ describes the VEVs of the boundary theory's single-trace operators (i.e. spin-$s$ conserved currents), while the magnetic data $A_{i_1\dots i_s}(\mathbf{x})$ describes the sources for these operators (i.e. spin-$s$ gauge fields).

Outside of transverse traceless gauge, it is helpful to characterize the two branches of solutions in a gauge-invariant way. For this purpose, we consider not the gauge potential $h_{\mu_1\dots\mu_s}$, but its field strength, the generalized Weyl tensor $\varphi_{\mu_1\nu_1\dots\mu_k\nu_k}$ from eq. \eqref{eq:C}. At every point, its linearly independent components are captured by a pair of totally symmetric traceless tensors $E_{i_1\dots i_s}$ and $B_{i_1\dots i_s}$, which describe respectively the field strength's electric and magnetic parts:
\begin{align}
 E_{i_1 i_2\dots i_s} = \varphi_{0 i_1 0 i_2\dots 0 i_s} \ ; \quad B_{i_1 i_2\dots i_s} = \frac{1}{2}\epsilon_{i_1 jk} \varphi^{jk}{}_{0 i_2\dots 0 i_s} \ . \label{eq:E_B}
\end{align}
On-shell, both of these are divergence-free in the 3d sense, i.e. $\del_{i_1} E^{i_1 i_2\dots i_s} = \del_{i_1} B^{i_1 i_2\dots i_s} = 0$. In the boundary limit $z\rightarrow 0$, both $E_{i_1\dots i_s}$ and $B_{i_1\dots i_s}$ scale as $z^{s+1}$:
\begin{align}
 E_{i_1\dots i_s}(z,\mathbf{x}) = z^{s+1}\calE_{i_1\dots i_s}(\mathbf{x}) \ ; \quad B_{i_1\dots i_s}(z,\mathbf{x}) = z^{s+1}\calB_{i_1\dots i_s}(\mathbf{x}) \ . \label{eq:E_B_boundary}
\end{align}
The magnetic branch of solutions \eqref{eq:magnetic_branch} can now be characterized by vanishing electric boundary data $\calE_{i_1\dots i_s}(\mathbf{x}) = 0$, while the electric branch \eqref{eq:electric_branch} is characterized by vanishing magnetic boundary data $\calB_{i_1\dots i_s}(\mathbf{x}) = 0$. Perhaps the quickest way to see this is to notice the behavior of the different types of boundary data under the antipodal map $z\rightarrow -z$: $A_{i_1\dots i_s}(\mathbf{x})$ and $\calB_{i_1\dots i_s}(\mathbf{x})$ are associated with antipodally even solutions, while $J_{i_1\dots i_s}(\mathbf{x})$ and $\calE_{i_1\dots i_s}(\mathbf{x})$ are associated with antipodally odd ones \cite{Neiman:2014npa,Halpern:2015zia} (in fact, $J_{i_1\dots i_s}(\mathbf{x})$ and $\calE_{i_1\dots i_s}(\mathbf{x})$ are the same up to a numerical factor, while $\calB_{i_1\dots i_s}(\mathbf{x})$ is a conformal field strength for the 3d gauge potential $A_{i_1\dots i_s}(\mathbf{x})$).

Returning now to transverse traceless gauge, let us work out the on-shell action of a free HS field in $EAdS_4$. A detailed analysis can be found in \cite{Joung:2011xb}; here, we will take some shortcuts towards the final answer. Since we are dealing with free field theory, and since $A_{i_1\dots i_s}(\mathbf{x})$ and $J_{i_1\dots i_s}(\mathbf{x})$ are canonically conjugate, the action (with divergent pieces removed) should be proportional to $\int A_{i_1\dots i_s}(\mathbf{x})\,J^{i_1\dots i_s}(\mathbf{x})\,d^3\mathbf{x}$. As we will now show, the correct proportionality factor is:
\begin{align}
 S[h,h] = \frac{1-2s}{2} \int A_{i_1\dots i_s}(\mathbf{x})\,J^{i_1\dots i_s}(\mathbf{x})\,d^3\mathbf{x} \ , \label{eq:S_EAdS}
\end{align}
where the notation on the LHS is intended to emphasize that $S$ is a quadratic form in the space of free-field solutions $h_{\mu_1\dots\mu_s}(z,\mathbf{x})$. The numerical coefficient in \eqref{eq:S_EAdS} depends on our choice of boundary terms for the action, which in turn depend on our choice of boundary conditions in the variational principle. Here, we are interested in the standard variational principle for AdS/CFT, in which the source-type boundary data $A_{i_1\dots i_s}(\mathbf{x})$ is held fixed. The variation of the action \eqref{eq:S_EAdS} then reads:
\begin{align}
 \delta S = (1-2s) \int J^{i_1\dots i_s}(\mathbf{x})\,\delta A_{i_1\dots i_s}(\mathbf{x})\,d^3\mathbf{x} \ . \label{eq:delta_S}
\end{align}
Identifying this as a symplectic potential, we take another variation to extract the symplectic form:
\begin{align}
 \Omega = (1-2s) \int \delta J^{i_1\dots i_s}(\mathbf{x})\wedge\delta A_{i_1\dots i_s}(\mathbf{x})\,d^3\mathbf{x} \ . \label{eq:Omega}
\end{align}
Note that in eq. \eqref{eq:S_EAdS}, $A_{i_1\dots i_s}(\mathbf{x})$ and $J^{i_1\dots i_s}(\mathbf{x})$ were linearly related by the requirement of regularity in $EAdS_4$, while in \eqref{eq:Omega}, we treat them as linearly independent.

We can now justify the numerical coefficient in \eqref{eq:S_EAdS} by comparing the symplectic form \eqref{eq:Omega} with the one derived directly from the Lagrangian $-\frac{1}{2}h_{\mu_1\dots\mu_s}G^{\mu_1\dots\mu_s}$. In general, this will be somewhat complicated, due to the large number of terms with different index arrangements inside $G^{\mu_1\dots\mu_s}$. However, in the boundary limit $z\rightarrow 0$ in transverse traceless gauge, things simplify considerably, and all the terms in $G^{\mu_1\dots\mu_s}$ beyond the ``trivial'' one $\Box h^{\mu_1\dots\mu_s}$ can be ignored. We then get the symplectic form:
\begin{align}
 \Omega =  \frac{1}{z^2} \int \delta h_{i_1\dots i_s}(z,\mathbf{x})\wedge\del_z\delta h^{i_1\dots i_s}(z,\mathbf{x})\,d^3\mathbf{x} \ , \label{eq:Omega_phi}
\end{align}
where the integral is at a fixed small value of $z$. By linear superposition of \eqref{eq:magnetic_branch} and \eqref{eq:electric_branch}, the general boundary behavior of $h_{i_1\dots i_s}$ is given by:
\begin{align}
 h_{i_1\dots i_s}(z,\mathbf{x}) = z^{2-s} A_{i_1\dots i_s}(\mathbf{x}) + z^{s+1} J_{i_1\dots i_s}(\mathbf{x}) + \dots \ , \label{eq:boundary_phi}
\end{align}
where the dots signify terms with $A_{i_1\dots i_s}(\mathbf{x})$ and its $\del_i$ derivatives multiplied by powers of $z$ higher than $z^{2-s}$, and terms with $J_{i_1\dots i_s}(\mathbf{x})$ and its $\del_i$ derivatives multiplied by powers of $z$ higher than $z^{s+1}$. In the symplectic form \eqref{eq:Omega_phi}, only the terms explicitly written in \eqref{eq:boundary_phi} will contribute. The other terms will end up vanishing, either due to high powers of $z$, or due to the wedge product's antisymmetry along with integration over the boundary. All in all, we see that the symplectic form \eqref{eq:Omega_phi} ends up conciding with \eqref{eq:Omega}, with the numerical coefficient arising as:
\begin{align}
 1-2s = \frac{1}{z^2}\left(z^{s+1}\del_z z^{2-s} - z^{2-s}\del_z z^{s+1} \right) \ .
\end{align}

\subsection{Embedding space, boundary-to-bulk propagators and 2-point function} \label{sec:Fronsdal:propagators}

While Poincare coordinates in $EAdS_4$ are sometimes useful, we prefer the more covariant embedding-space picture. There, $EAdS_4$ is defined as the hyperboloid of unit timelike radius embedded in a flat Minkowski space $\bbR^{1,4}$:
\begin{align}
 EAdS_4 = \left\{x^\mu\in\bbR^{1,4}\, |\, x_\mu x^\mu = -1, \ x^0 > 0 \right\} \ . \label{eq:EAdS}
\end{align}
The flat metric $\eta_{\mu\nu}$ of $\bbR^{1,4}$ has mostly-plus signature. In an abuse of notation, we will denote the 5d embedding-space indices by the same Greek letters $(\mu,\nu,\dots)$ that we used for intrinsic tensors in the 4d spacetime. This is quite natural, because intrinsic $EAdS_4$ vectors at a point $x^\mu\in EAdS_4$ are simply vectors $v^\mu\in\bbR^{1,4}$ that happen to be tangent to the hyperboloid \eqref{eq:EAdS}, i.e. that satisfy $x\cdot v \equiv x_\mu v^\mu = 0$. In particular, the $EAdS_4$ metric is simply $g_{\mu\nu}(x) = \eta_{\mu\nu} + x_\mu x_\nu$. Similarly, the $EAdS_4$ covariant derivative $\nabla_\mu$ is just the flat $\bbR^{1,4}$ derivative $\del_\mu$, followed by projecting all tensor indices back into the hyperboloid, using the projector $\delta_\mu^\nu + x_\mu x^\nu$. With this notation, the formulas of section \ref{sec:Fronsdal:equations} carry through seamlessly.

The conformal boundary of $EAdS_4$ is defined by the projective lightcone in the $\bbR^{1,4}$ embedding space, i.e. by null vectors $\ell^\mu\in\bbR^{1,4}$, $\ell\cdot\ell = 0$, modulo equivalence under rescalings $\ell^\mu \rightarrow \rho\ell^\mu$. The bulk$\rightarrow$boundary limit can be described as $x^\mu\rightarrow \ell^\mu/z$, where the parameter $z$ goes to zero (as the coincident notation suggests, the Poincare coordinate $z$ near the boundary plays this role). Vectors on the 3d conformal boundary are described by $\bbR^{1,4}$ vectors $v^\mu$ that are tangential to the lightcone, $v\cdot\ell = 0$, subject to the equivalence relation $v^\mu \cong v^\mu + \alpha\ell^\mu$. Since our axes in $\bbR^{1,4}$ are orthonormal, many of the formulas from section \ref{sec:Fronsdal:boundary} carry over to this picture, with embedding-space indices $(\mu,\nu,\dots)$ in place of the orthonormal boundary indices $(i,j,\dots)$.

Let us now use the embedding-space language to write down massless spin-$s$ boundary-to-bulk propagators. These are solutions to the free bulk field equations that are characterized by a ``source'' point $\ell^\mu$ on the boundary, and a null polarization vector $\lambda^\mu$ at that point, so that $\ell\cdot\ell = \ell\cdot\lambda = \lambda\cdot\lambda = 0$. The equivalence relation $\lambda^\mu \cong \lambda^\mu + \alpha\ell^\mu$ can be made manifest by replacing $\lambda^\mu$ with the totally-null bivector $M^{\mu\nu} \equiv 2\ell^{[\mu}\lambda^{\nu]}$, which satisfies:
\begin{align}
 M_{\mu\nu}\ell^\nu = M_{\mu\nu}M^{\nu\rho} = M_{[\mu\nu}M_{\rho]\sigma} = 0 \ .
\end{align}
At a bulk point $x^\mu$, the boundary inputs $(\ell^\mu,M^{\mu\nu})$ induce three quantities: a scalar $\ell\cdot x$ and two vectors $\ell^\mu_\perp \equiv \ell^\mu + (\ell\cdot x)x^\mu$ and $m^\mu\equiv M^{\mu\nu}x_\nu$, which satisfy:
\begin{align}
 m_\mu\ell_\perp^\mu = m_\mu m^\mu = 0 \ ; \quad \ell_\mu^\perp = \nabla_\mu(\ell\cdot x) \ .
\end{align}
The ``$\perp$'' label on $\ell^\mu_\perp$ indicates projection in perpendicular to $x^\mu$, i.e. into the $EAdS_4$ tangent space at $x$. The projection of $M^{\mu\nu}$ into the tangent space at $x$ reads:
\begin{align}
 M_\perp^{\mu\nu} \equiv M^{\mu\nu} + 2m^{[\mu}x^{\nu]} = \frac{2m^{[\mu}\ell_\perp^{\nu]}}{\ell\cdot x} \ . \label{eq:M_perp}
\end{align}
The spin-$s$ boundary-to-bulk propagator is now given by:
\begin{align}
 h_{\mu_1\dots\mu_s}(x) = \frac{M_{\mu_1\nu_1}x^{\nu_1}\dots M_{\mu_s\nu_s}x^{\nu_s}}{(\ell\cdot x)^{2s+1}} = \frac{m_{\mu_1}\dots m_{\mu_s}}{(\ell\cdot x)^{2s+1}} \ . \label{eq:boundary_bulk}
\end{align}
It's easy to verify that this satisfies Fronsdal's equations \eqref{eq:TT_eqs} in transverse traceless gauge. The field strength \eqref{eq:C} associated with this solution can be calculated by repeatedly applying the identities:
\begin{align}
  m_{[\mu}\nabla_{\nu]}(\ell\cdot x) = \frac{1}{2}(\ell\cdot x)M^\perp_{\mu\nu} \ ; \quad \nabla_\mu m_\nu = -M^\perp_{\mu\nu} \ ; \quad m_{[\mu}\nabla_{\nu]}m_\rho = \frac{1}{2}M^\perp_{\mu\nu}m_\rho \ ,
\end{align}
and noting that $\nabla_\mu M^\perp_{\nu\rho} = -2g_{\mu[\nu}m_{\rho]}$ can be discarded as a trace piece. The result reads:
\begin{align}
 \varphi_{\mu_1\nu_1\dots\mu_s\nu_s}(x) = \frac{(2s-1)!}{(s-1)!} \cdot \frac{M^\perp_{\mu_1\nu_1}\dots M^\perp_{\mu_s\nu_s}}{(\ell\cdot x)^{2s+1}} - \text{traces} \ . \label{eq:boundary_bulk_field_strength_raw}
\end{align}
The subtraction of traces is equivalent to leaving just the purely right-handed and purely left-handed parts of $M^\perp_{\mu_1\nu_1}\dots M^\perp_{\mu_s\nu_s}$. Thus, if we define the left/right-handed parts of $M^\perp_{\mu\nu}$ as:
\begin{align}
 M^{L/R}_{\mu\nu} \equiv \frac{1}{2}\left(M^\perp_{\mu\nu} \pm \frac{1}{2}\epsilon_{\mu\nu}{}^{\lambda\rho\sigma}x_\lambda M_{\rho\sigma}\right) \ , \label{eq:M_chiral}
\end{align}
then the field strength \eqref{eq:boundary_bulk_field_strength_raw} can be written as:
\begin{align}
 s = 0:\quad &\varphi(x) = \frac{1}{\ell\cdot x} \ ; \label{eq:boundary_bulk_field_strength_0} \\
 s\geq 1:\quad &\varphi_{\mu_1\nu_1\dots\mu_s\nu_s}(x) = \frac{(2s-1)!}{(s-1)!} \cdot \frac{M^L_{\mu_1\nu_1}\dots M^L_{\mu_s\nu_s} + M^R_{\mu_1\nu_1}\dots M^R_{\mu_s\nu_s}}{(\ell\cdot x)^{2s+1}} \ , \label{eq:boundary_bulk_field_strength_s}
\end{align}
where we included the spin-0 case separately. Note that for $s=0$, the factorials in \eqref{eq:boundary_bulk_field_strength_raw},\eqref{eq:boundary_bulk_field_strength_s} become ill-defined. However, we can analytically continue to continuous values of $s$ (whereupon the factorials become Gamma functions), and then take the limit $s\rightarrow 0$. We then see that \eqref{eq:boundary_bulk_field_strength_raw} (but not \eqref{eq:boundary_bulk_field_strength_s}) correctly reproduces the spin-0 case $\varphi(x) = h(x) = 1/(\ell\cdot x)$.

Let's now identify the asymptotic behavior of the boundary-to-bulk propagator \eqref{eq:boundary_bulk}. At a boundary point $\hat\ell\neq\ell$, the boundary data is purely electric, and can be read off directly from eq. \eqref{eq:boundary_bulk} as:
\begin{align}
 J_{\mu_1\dots\mu_s}(\hat\ell) = \frac{M_{\mu_1\nu_1}\hat\ell^{\nu_1}\dots M_{\mu_s\nu_s}\hat\ell^{\nu_s}}{(\ell\cdot\hat\ell)^{2s+1}} \ . \label{eq:local_J}
\end{align}
The magnetic boundary data for the propagator \eqref{eq:boundary_bulk} takes the form of a delta function supported at $\hat\ell = \ell$:
\begin{align}
 s &= 0: & A(\hat\ell) &= 4\pi^2 \delta^3(\hat\ell,\ell) \ ; \label{eq:A_0} \\
 s&\geq 1: & A_{\mu_1\dots\mu_s}(\hat\ell) &= -\frac{4\pi^2(2s-2)!}{2^s s!(s-1)!}\,\delta^3(\hat\ell,\ell)\,\lambda_{\mu_1}\!\dots\lambda_{\mu_s} \ . \label{eq:A_s}
\end{align}
The numerical coefficient in \eqref{eq:A_s} is rather non-trivial to derive, and we won't reproduce the derivation here. It's been worked out, with small mistakes, in \cite{Mikhailov:2002bp}, as well as by one of the present authors in \cite{Halpern:2015zia}. A correct derivation can now be found in the updated version of \cite{Halpern:2015zia}. As with eq. \eqref{eq:boundary_bulk_field_strength_raw} above, though the coefficient in \eqref{eq:A_s} is ill-defined for $s=0$, we can reproduce the spin-0 case \eqref{eq:A_0} by making $s$ continuous and then taking the limit $s\rightarrow 0$.

Finally, we can plug the boundary data \eqref{eq:local_J}-\eqref{eq:A_s} into the action formula \eqref{eq:S_EAdS} to obtain the 2-point function. Specifically, if we consider two boundary-to-bulk propagators of the form \eqref{eq:boundary_bulk} with parameters $(\ell_1^\mu,M_1^{\mu\nu} = 2\ell_1^{[\mu}\lambda_1^{\nu]})$ and $(\ell_2^\mu,M_2^{\mu\nu})$, then their contribution to the quadratic action \eqref{eq:S_EAdS} reads:
\begin{align}
 S[h_1,h_2] &= \frac{1}{2}\int d^3\hat\ell\, A_1(\hat\ell) J_2(\hat\ell) = 2\pi^2 J_2(\ell_1) = \frac{2\pi^2}{\ell_1\cdot\ell_2} \label{eq:S_local_0}
\end{align}
for $s=0$, and:
\begin{align}
 \begin{split}
   S[h_1,h_2] &= \frac{1-2s}{2}\int d^3\hat\ell\, A^1_{\mu_1\dots\mu_s}(\hat\ell) J_2^{\mu_1\dots\mu_s}(\hat\ell) = \frac{2\pi^2(2s-1)!}{2^s s!(s-1)!}\,\lambda_{\mu_1}\!\dots\lambda_{\mu_s} J_2^{\mu_1\dots\mu_s}(\ell_1) \\
    &= \frac{(-1)^s\pi^2(2s)!}{4^s (s!)^2}\cdot\frac{(M^1_{\mu\nu} M_2^{\mu\nu})^s}{(\ell_1\cdot\ell_2)^{2s+1}} \ 
 \end{split} \label{eq:S_local_s}
\end{align}
for $s\geq 1$. Again, \eqref{eq:S_local_0} can be regarded a special case of \eqref{eq:S_local_s}, by analytically continuing to continuous $s$ and then sending $s\rightarrow 0$.

\section{HS fields with bulk particle sources} \label{sec:bulk_particle}

In this section, we couple Fronsdal's linearized HS fields in $\bbR^4$ and $EAdS_4$ to a particle-like source, with support on a geodesic worldline $\gamma$. 

\subsection{Action and field equations}

First, consider coupling the spin-$s$ Fronsdal field to a general HS current $T^{\mu_1\dots\mu_s}$:
\begin{align}
 S = \int d^4x\sqrt{g}\left(-\frac{1}{2}h_{\mu_1\dots\mu_s}G^{\mu_1\dots\mu_s} + h_{\mu_1\dots\mu_s}T^{\mu_1\dots\mu_s} \right) +  \text{boundary terms} \ . \label{eq:S_with_current}
\end{align}
The current $T^{\mu_1\dots\mu_s}$ inherits the algebraic properties of $ h_{\mu_1\dots\mu_s}$, i.e. we take it to be totally symmetric and double-traceless. In addition, invariance under the gauge symmetry \eqref{eq:gauge_transformation} demands that $T^{\mu_1\dots\mu_s}$ be conserved in the same sense as $G^{\mu_1\dots\mu_s}$, i.e. that the \emph{traceless part} of $\nabla_\nu T^\nu_{\mu_1\dots\mu_{s-1}}$ should vanish:
\begin{align}
 \nabla_\nu T^\nu_{\mu_1\dots\mu_{s-1}} = \text{trace terms only} \ . \label{eq:T_conservation}
\end{align}
Varying the action \eqref{eq:S_with_current} with respect to $h_{\mu_1\dots\mu_s}$, we obtain the field equations $G^{\mu_1\dots\mu_s} = T^{\mu_1\dots\mu_s}$. Rearranging the trace as in \eqref{eq:F_G}, we express these equations in terms of the Fronsdal tensor:
\begin{align}
 F^{\mu_1\dots\mu_s} = T^{\mu_1\dots\mu_s} - \frac{s}{4}\,g^{(\mu_1\mu_2} T_\nu^{\mu_3\dots\mu_s)\nu} \ . \label{eq:field_eqs_with_current}
\end{align}
Now, what HS current $T^{\mu_1\dots\mu_s}$ can we associate with a point particle? Our first building block is the delta function $\int_\gamma d\tau\,\delta^4(x,x')$ that localizes the particle on its worldline $\gamma$; here, $x^\mu$ are the worldline's coordinates, and $d\tau = \sqrt{dx_\mu dx^\mu}$ is the length element. The second bulding block is the particle's 4-velocity $u^\mu = dx^\mu/d\tau$. We will assume minimal coupling, which forbids spacetime derivatives of the delta function. We further assume that the worldline is a geodesic, so that any further $\tau$ derivatives of $u^\mu$ vanish. The most general totally symmetric tensor then reads:
\begin{align}
 T^{\mu_1\dots\mu_s}(x') = \int_\gamma d\tau\,\delta^4(x,x') \sum_{n=0}^{\lfloor s/2 \rfloor} Q^{(s)}_n g^{(\mu_1\mu_2}\dots g^{\mu_{2n-1}\mu_{2n}} u^{\mu_{2n+1}}\dots u^{\mu_s)} \ , \label{eq:T_raw}
\end{align}
with some coefficients $Q^{(s)}_n$. The $n=0$ term in \eqref{eq:T_raw} is conserved, thanks to the geodesic condition $u^\nu\nabla_\nu u^\mu = 0$. The $n\geq 2$ terms are double-trace pieces, which are fixed by the requirement that $T^{\mu_1\dots\mu_s}$ is double-traceless overall. Their divergence is automatically a trace piece, so they do not affect the conservation condition \eqref{eq:T_conservation}. This leaves the $n=1$ term, which does \emph{not} satisfy the conservation law \eqref{eq:T_conservation}, and must therefore be ruled out. All in all, we end up with the simplest possible coupling between the spin-$s$ field and a point particle:
\begin{align}
&T^{\mu_1\dots\mu_s}(x') = Q^{(s)}\!\int_\gamma d\tau\,\delta^4(x,x')\,u^{\mu_1}\!\dots u^{\mu_s} - \text{double traces} \ ; \label{eq:T} \\
&S = -\frac{1}{2}\int d^4x\sqrt{g}\,h_{\mu_1\dots\mu_s}G^{\mu_1\dots\mu_s} + Q^{(s)}\!\int_\gamma d\tau\,h_{\mu_1\dots\mu_s}u^{\mu_1}\!\dots u^{\mu_s} +  \text{boundary terms} \ , \label{eq:S_with_particle}
\end{align}
where we renamed $Q^{(s)}_0\equiv Q^{(s)}$ for brevity. The field equations \eqref{eq:field_eqs_with_current} in the presence of the current \eqref{eq:T} read:
\begin{align}
 F^{\mu_1\dots\mu_s}(x') = Q^{(s)}\!\int_\gamma d\tau\,\delta^4(x,x') \left(u^{\mu_1}\!\dots u^{\mu_s} - \frac{s}{4}\,g^{(\mu_1\mu_2}u^{\mu_3}\!\dots u^{\mu_s)}\right) - \text{double traces} \ . \label{eq:field_eqs_with_particle}
\end{align}

\subsection{Solution of the field equations in $\bbR^4$} \label{sec:bulk_particle:flat}

Let us now solve the field equations \eqref{eq:field_eqs_with_particle}. We begin with the $\bbR^4$ case. The worldline $\gamma$ is now just a straight line, which we can think of as running along the (Euclidean) time direction. Let us extend the 4-velocity $u^\mu$ into a constant vector field in spacetime, which we'll denote as $t^\mu$. Let $R$ denote our distance from the worldline, and let $r^\mu$ be the corresponding radius-vector (we denote these by different-case letters, for consistency with the curved case below). We then have the basic identities:
\begin{align}
 \begin{split}
   t_\mu t^\mu &= 1 \ ; \quad r_\mu r^\mu = R^2 \ ; \quad t_\mu r^\mu = 0 \ ; \\
   \nabla_\mu t_\nu &= 0 \ ; \quad \nabla_\mu R = \frac{r_\mu}{R} \ ; \quad \nabla_\mu r_\nu = q_{\mu\nu} \ ,
 \end{split} \label{eq:t_r_flat}
\end{align}
where $q_{\mu\nu} = g_{\mu\nu} - t_\mu t_\nu$ is the flat metric of the 3d space orthogonal to $t^\mu$. 

\subsubsection{The solution}

For spin 0, the field equation \eqref{eq:field_eqs_with_particle} and its solution read:
\begin{align}
 \Box h = Q^{(0)}\delta^3(r) \ ; \quad h(x) = -\frac{Q^{(0)}}{4\pi R} \ . \label{eq:flat_solution_0}
\end{align} 
For nonzero spins, we will work in a gauge that is traceless (but not transverse). The field equation \eqref{eq:field_eqs_with_particle} then reads:
\begin{align}
 \Box h_{\mu_1\dots\mu_s} - s\nabla_{(\mu_1}\nabla^\nu h_{\mu_2\dots\mu_s)\nu} = Q^{(s)}\delta^3(r)\left(t^{\mu_1}\!\dots t^{\mu_s} - \frac{s}{4}\,g^{(\mu_1\mu_2}t^{\mu_3}\!\dots t^{\mu_s)} - \text{double traces} \right) \ . \label{eq:field_eqs_flat}
\end{align}
To solve it, we define a null combination of $t^\mu$ and $r^\mu$:
\begin{align}
 k^\mu \equiv \frac{1}{2}\left(t^\mu + \frac{ir^\mu}{R}\right) \ . \label{eq:k}
\end{align}
In Lorentzian signature, this would be a lightlike vector, which defines an affine tangent to the lightrays emanating from the worldline. In terms of $k^\mu$ and $r^\mu$, eqs. \eqref{eq:t_r_flat} become:
\begin{align}
 \begin{split}
   k_\mu k^\mu &= 0 \ ; \quad r_\mu r^\mu = R^2 \ ; \quad k_\mu r^\mu = \frac{iR}{2} \ ; \\
   \nabla_\mu k_\nu &= \frac{i\Omega_{\mu\nu}}{2R} \ ; \quad \nabla_\mu R = \frac{r_\mu}{R} \ ; \quad \nabla_\mu r_\nu = q_{\mu\nu} \ , 
 \end{split} \label{eq:k_r_flat}
\end{align}
where $\Omega_{\mu\nu} = q_{\mu\nu} - \frac{1}{R^2}r_\mu r_\nu$ is the metric of the 2-sphere at radius $r$. In terms of $k^\mu$ and $r^\mu$, the metrics $q_{\mu\nu}$ and $\Omega_{\mu\nu}$ take the form:
\begin{align}
 \Omega_{\mu\nu} = g_{\mu\nu} - 4k_\mu k_\nu + \frac{4i}{R}k_{(\mu}r_{\nu)} \ ; \quad q_{\mu\nu} = \Omega_{\mu\nu} + \frac{r_\mu r_\nu}{R^2} \ . \label{eq:Omega_q_flat}
\end{align}
We now claim that the following Kerr-Schild-like field, familiar from \cite{Didenko:2008va}, solves the field equation \eqref{eq:field_eqs_flat} for all nonzero spins $s\geq 1$:
\begin{align}
 h_{\mu_1\dots\mu_s}(x) = -\frac{Q^{(s)}}{2\pi R}\,k_{\mu_1}\!\dots k_{\mu_s} \ . \label{eq:flat_solution_s}
\end{align}
Note that this differs by a factor of 2 from the $s=0$ case \eqref{eq:flat_solution_0}. The solution \eqref{eq:flat_solution_s} is traceless as promised, since $k^\mu$ is null. Plugging it into the field equation \eqref{eq:field_eqs_flat}, one can easily verify that the LHS vanishes at $R\neq 0$, using the following corrolaries of eqs. \eqref{eq:k_r_flat}-\eqref{eq:Omega_q_flat}:
\begin{align}
 \begin{split}
   &\nabla_\mu k^\mu = \frac{i}{R} \ ; \quad k^\nu\nabla_\nu k_\mu = r^\nu\nabla_\nu k_\mu = 0 \ ; \\
   &\Box \frac{1}{R} = 0 \ ; \quad \Box k_\mu = -\frac{ir_\mu}{R^3} \ ; \quad \nabla^\rho k_\mu \nabla_\rho k_\nu = -\frac{1}{R^2}\left(\frac{1}{4}g_{\mu\nu} - k_\mu k_\nu + \frac{ik_{(\mu}r_{\nu)}}{R} \right) \ .
 \end{split} \label{eq:box_flat}
\end{align}
What remains is to resolve the delta-function-like source at $R=0$. For that purpose, we write the Fronsdal tensor on the LHS of \eqref{eq:field_eqs_flat} as a total divergence:
\begin{align}
 F_{\mu_1\dots\mu_s} = \nabla_\nu K^\nu{}_{\mu_1\dots\mu_s} \ ; \quad K^\nu{}_{\mu_1\dots\mu_s} = \nabla^\nu h_{\mu_1\dots\mu_s} - s\nabla_{(\mu_1}h^\nu_{\mu_2\dots\mu_s)} \ .
\end{align}
We now need to show that the flux of $K^\nu{}_{\mu_1\dots\mu_s}$ through a 2-sphere at radius $R$ reproduces the coefficient of the delta function on the RHS of \eqref{eq:field_eqs_flat}:
\begin{align}
  4\pi R \langle r_\nu K^\nu{}_{\mu_1\dots\mu_s} \rangle_{S_2} = Q^{(s)}\left( t^{\mu_1}\!\dots t^{\mu_s} - \frac{s}{4}\,g^{(\mu_1\mu_2}t^{\mu_3}\!\dots t^{\mu_s)} - \text{double traces}\right) \ . \label{eq:delta_source}
\end{align}
As a first step, we note that $r_\nu K^\nu{}_{\mu_1\dots\mu_s}$ is double-traceless already before the $S_2$ averaging, thanks to the tracelessness of $h_{\mu_1\dots\mu_s}$. When evaluated explicitly, it reads:
\begin{align}
 \begin{split}
   r_\nu K^\nu{}_{\mu_1\dots\mu_s} = \frac{Q^{(s)}}{2\pi R}&\left(k_{\mu_1}\!\dots k_{\mu_s} - \frac{is}{2R}\,r_{(\mu_1}k_{\mu_2}\!\dots k_{\mu_s)} - \frac{s(s-1)}{4}\,\Omega_{(\mu_1\mu_2}k_{\mu_3}\!\dots k_{\mu_s)} \right) \\ 
     = \frac{Q^{(s)}}{2\pi R}&\left(k_{\mu_1}\!\dots k_{\mu_s} - \frac{is}{2R}\,r_{(\mu_1}k_{\mu_2}\!\dots k_{\mu_s)} + \frac{s(s-1)}{4R^2}\,r_{(\mu_1}r_{\mu_2}k_{\mu_3}\!\dots k_{\mu_s)} \right. \\
       &\left.\quad {}- \frac{s(s-1)}{4}\,q_{(\mu_1\mu_2}k_{\mu_3}\!\dots k_{\mu_s)} \right) \ .
 \end{split} \label{eq:rK_raw}
\end{align}
The double-tracelessness is manifest in the first line, since $k_\mu$ is null and orthogonal to $\Omega_{\mu\nu}$. To perform the $S_2$ average, we decompose \eqref{eq:rK_raw} along $t_\mu$, $q_{\mu\nu}$ and $r_\mu$:
\begin{align}
 \begin{split}
   r_\nu K^\nu{}_{\mu_1\dots\mu_s} ={}& \frac{s!\,Q^{(s)}}{2^{s+1}\pi}\left(\sum_{n=0}^s \frac{i^n (1-n^2) r_{(\mu_1}\!\dots r_{\mu_n} t_{\mu_{n+1}}\!\dots t_{\mu_s)} }{n!(s-n)!R^{n+1}} \right. \\
    &-\left. \sum_{n=0}^{s-2} \frac{i^n r_{(\mu_1}\!\dots r_{\mu_n}q_{\mu_{n+1}\mu_{n+2}}t_{\mu_{n+3}}\!\dots t_{\mu_s)}}{n!(s-n-2)!R^{n+1}} \right) \ .
 \end{split} \label{eq:rK}
\end{align}
The $S_2$ averaging now affects only the $r_{\mu_1}\!\dots r_{\mu_n}$ factors. For odd $n$, these average to zero, while for even $n$, we have the identity:
\begin{align}
 \langle r_{\mu_1}\!\dots r_{\mu_n} \rangle_{S_2} = \frac{R^n}{n+1}\,q_{(\mu_1\mu_2}\dots q_{\mu_{n-1}\mu_n)} \ . \label{eq:r_n_average}
\end{align}
The latter can be proved by contracting both sides with $q^{\mu_1\mu_2}\dots q^{\mu_{n-1}\mu_n}$, and using the identity:
\begin{align}
 q^{\mu_1\mu_2}\dots q^{\mu_{n-1}\mu_n}q_{(\mu_1\mu_2}\dots q_{\mu_{n-1}\mu_n)} = n+1 \ ,
\end{align}
which is easy to prove recursively in $n$. Plugging \eqref{eq:r_n_average} into \eqref{eq:rK}, we find that the two terms in \eqref{eq:rK} combine nicely, giving:
\begin{align}
 \langle r_\nu K^\nu{}_{\mu_1\dots\mu_s} \rangle_{S_2} = \frac{Q^{(s)}}{2^{s+1}\pi R}\sum_{n=0}^{\lfloor s/2 \rfloor} (-1)^n \binom{s}{2n} q_{(\mu_1\mu_2}\dots q_{\mu_{2n-1}\mu_{2n}} t_{\mu_{2n+1}}\!\dots t_{\mu_s)} \ . \label{eq:rK_average}
\end{align}
To compare with \eqref{eq:delta_source}, we must re-express the sum in \eqref{eq:rK_average} in terms of $g_{\mu\nu}$ and $t_\mu$, by substituting $q_{\mu\nu} = g_{\mu\nu} - t_\mu t_\nu$. Since the double-tracelessness is assured, it's enough to compare the coefficients of $t_{\mu_1}\!\dots t_{\mu_s}$ and of $g_{(\mu_1\mu_2}t_{\mu_3}\!\dots t_{\mu_s)}$. This is easy to do, confirming the flux relation \eqref{eq:delta_source}, and with it the field equation \eqref{eq:field_eqs_flat}.

\subsubsection{Field strength and symmetric gauge}

We can now calculate the Weyl-like field strength \eqref{eq:C} of the solution \eqref{eq:flat_solution_s}:
\begin{align}
 \varphi_{\mu_1\nu_1\dots\mu_s\nu_s}(x) = - \frac{(2s)!}{s!}\cdot\frac{Q^{(s)} S^\perp_{\mu_1\nu_1}\dots S^\perp_{\mu_s\nu_s}}{4\pi R^{2s+1}} - \text{traces} \ , \label{eq:C_flat_raw}
\end{align}
where $S^\perp_{\mu\nu}$ is a bivector in the $tr$ plane:
\begin{align}
 S^\perp_{\mu\nu} \equiv t_{[\mu}r_{\nu]} = 2k_{[\mu}r_{\nu]} \ , \label{eq:flat_bivector}
\end{align}
and the ``$\perp$'' superscript is in anticipation of the $EAdS_4$ case below. The derivation of the field strength \eqref{eq:C_flat_raw} from the potential \eqref{eq:flat_solution_s} is easy, once it is organized in terms of $S^\perp_{\mu\nu}$. The relevant identities read:
\begin{align}
 k_{[\mu}\nabla_{\nu]} k_\rho = \frac{S^\perp_{\mu\nu} k_\rho}{2R^2} + \frac{i}{2R}k_{[\mu}g_{\nu]\rho} \ ; \quad \nabla_\mu S^\perp_{\nu\rho} = t_{[\nu}g_{\rho]\mu} \ , \label{eq:flat_curls}
\end{align}
where every term proportional to $g_{\mu\nu}$ can be discarded as a trace piece. Note that, unlike the potential \eqref{eq:flat_solution_s}, the field strength \eqref{eq:C_flat} correctly covers also the spin-0 case \eqref{eq:flat_solution_0}. The subtraction of traces in \eqref{eq:C_flat_raw} can again be expressed as a projection onto the purely right-handed and purely left-handed parts:
\begin{align}
 \varphi_{\mu_1\nu_1\dots\mu_s\nu_s}(x) = - \frac{(2s)!}{s!}\cdot\frac{Q^{(s)}}{4\pi R^{2s+1}}\left(S^L_{\mu_1\nu_1}\dots S^L_{\mu_s\nu_s} + S^R_{\mu_1\nu_1}\dots S^R_{\mu_s\nu_s}\right) \ , \label{eq:C_flat}
\end{align}
where:
\begin{align}
S^{R/L}_{\mu\nu} \equiv \frac{1}{2}\left(S^\perp_{\mu\nu} \pm \frac{1}{2}\epsilon_{\mu\nu}{}^{\rho\sigma} S^\perp_{\rho\sigma}\right) \ .
\end{align}
Finally, we note the freedom of gauge-transforming the solution \eqref{eq:flat_solution_s}. One reason to prefer a different gauge is that \eqref{eq:flat_solution_s} discriminates between the two null vectors $k_\mu = \frac{1}{2}(t_\mu + \frac{i}{R}r_\mu)$ and $\bar k_\mu = \frac{1}{2}(t_\mu - \frac{i}{R}r_\mu)$, thus breaking time-reversal symmetry $t\rightarrow -t$ (in the context of black holes, this can be natural, if one wishes to ignore the time-reversed white hole). It's easy to see that if we complex-conjugate the solution \eqref{eq:flat_solution_s}, i.e. replace $k_\mu\rightarrow \bar k_\mu$ everywhere, the Fronsdal tensor \eqref{eq:field_eqs_flat} and Weyl-like field strength \eqref{eq:C_flat}, which are real, remain unchanged. Therefore, replacing $k_\mu\rightarrow \bar k_\mu$ is a gauge transformation. Taking the average of \eqref{eq:flat_solution_s} and its complex conjugate, we obtain the solution in a gauge that is real and symmetric under time reversal:
\begin{align}
 h_{\mu_1\dots\mu_s}(x) = -\frac{Q^{(s)}}{4\pi R}\left(k_{\mu_1}\!\dots k_{\mu_s} + \bar k_{\mu_1}\!\dots\bar k_{\mu_s}\right) = -\frac{Q^{(s)}}{2\pi R}\Re(k_{\mu_1}\!\dots k_{\mu_s}) \ . \label{eq:flat_solution_symm}
\end{align}
For $s=1$, this is the usual Coulomb potential $-Q^{(1)} t_\mu/(4\pi R)$.

\subsection{Solution of the field equations in $EAdS_4$} \label{sec:bulk_particle:EAdS}

Now, consider the same field equation with a particle source \eqref{eq:field_eqs_with_particle} in $EAdS_4$, where we again work in the embedding-space formalism. The source particle's geodesic worldline now stretches from one point $\ell'^\mu$ to another $\ell^\mu$ on the $EAdS_4$ boundary (see figure \ref{fig:3pt}a). In terms of these, the worldline and its 4-velocity are given by:
\begin{figure}%
	\centering%
	\includegraphics[scale=0.6]{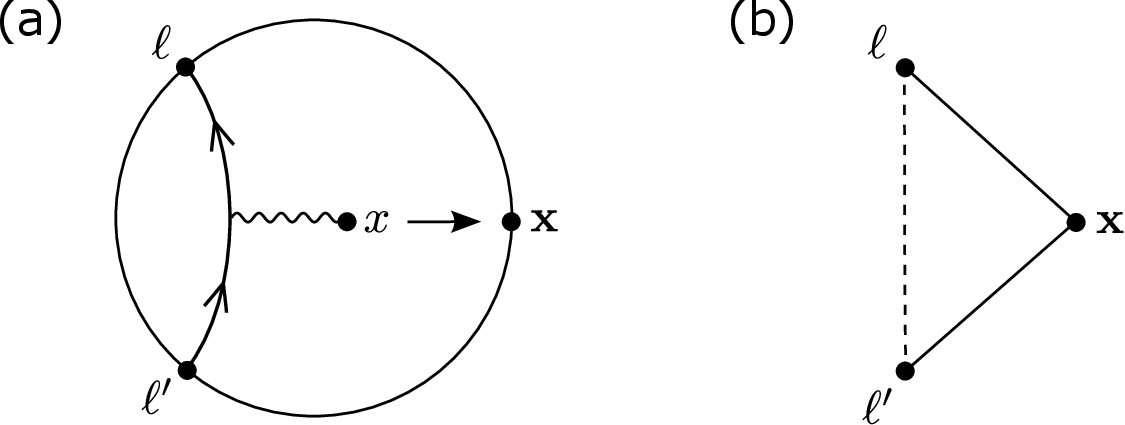} \\
	\caption{(a) An HS-charged particle traveling along a geodesic between two boundary points $\ell,\ell'$ in $EAdS_4$ is generating HS gauge fields at a bulk point $x$. (b) A Feynman diagram in the boundary vector model, connecting a bilocal operator $\bar\phi_I(\ell')\phi^I(\ell)$ to a local current at $\mathbf{x}$, which can be thought of as a boundary limit of the bulk point $x$; the solid lines are propagators, while the dashed line simply contracts the color index $I$. The boundary diagram in (b) can be viewed as an HS multiplet of OPE blocks, between the two fundamental fields $\phi^I(\ell),\bar\phi_I(\ell')$ and the currents $j^{(s)}(\mathbf{x})$ of all spins, which constitute the full OPE of $\phi^I(\ell)$ and $\bar\phi_I(\ell')$. The bulk picture in (a), with the particle assigned the DV pattern of charges, is a geodesic Witten diagram for these OPE blocks.}
	\label{fig:3pt} 
\end{figure}%
\begin{align}
 x^\mu(\tau) = \frac{1}{\sqrt{-2\ell\cdot\ell'}}\left(e^\tau\ell^\mu + e^{-\tau}\ell'^\mu \right) \ ; \quad u^\mu(\tau) = \frac{1}{\sqrt{-2\ell\cdot\ell'}}\left(e^\tau\ell^\mu - e^{-\tau}\ell'^\mu \right) \ .
\end{align}
To each spacetime point $x^\mu\in EAdS_4$ away from the worldline, we can again associate a radial parameter $R$, a radial vector $r^\mu$ and a ``time'' vector $t^\mu$:
\begin{align}
 R &= \sqrt{-\frac{2(x\cdot\ell)(x\cdot\ell')}{\ell\cdot\ell'} - 1} \ ; \label{eq:R} \\ 
 r^\mu &= x^\mu + \frac{1}{2}\left(\frac{\ell^\mu}{x\cdot\ell} + \frac{\ell'^\mu}{x\cdot\ell'} \right) \ ; \quad t^\mu = \frac{1}{2}\left(\frac{\ell'^\mu}{x\cdot\ell'} - \frac{\ell^\mu}{x\cdot\ell} \right) \ . \label{eq:r_t}
\end{align}
$r^\mu$ and $t^\mu$ lie in the $EAdS_4$ tangent space at $x$, i.e. $x_\mu r^\mu = x_\mu t^\mu = 0$. The variables $(R,r^\mu,t^\mu)$ play similar roles to their flat counterparts from section \ref{sec:bulk_particle:flat}, and coincide with them close to the worldline, where the curvature of $EAdS_4$ can be neglected. In particular, $R$ is a measure of distance from the worldline, $r^\mu$ is tangent to the radial geodesics that emanate from the worldline perpendicularly, and $t^\mu$ is an extension of the 4-velocity $u^\mu$ from the worldline into the rest of spacetime. The identities \eqref{eq:t_r_flat} acquire curvature corrections, as:
\begin{align}
 \begin{split}
  t_\mu t^\mu &= \frac{1}{1+R^2} \ ; \quad r_\mu r^\mu = \frac{R^2}{1+R^2} \ ; \quad t_\mu r^\mu = 0 \ ; \\
  \nabla_\mu t_\nu &= -2t_{(\mu}r_{\nu)} \ ; \quad \nabla_\mu R = \frac{1+R^2}{R}\,r_\mu \ ; \quad \nabla_\mu r_\nu = g_{\mu\nu} - t_\mu t_\nu - r_\mu r_\nu \ .
 \end{split} \label{eq:t_r_curved}
\end{align}
We again define a null combination $k^\mu$ of $t^\mu$ and $r^\mu$, according to eq. \eqref{eq:k}. In Lorentzian, this would again be an affine tangent to the lightrays emanating from the worldline. The identities \eqref{eq:k_r_flat}-\eqref{eq:box_flat} with curvature corrections read:
\begin{align}
 \begin{split}
   &k_\mu k^\mu = 0 \ ; \quad k_\mu r^\mu = \frac{iR}{2(1+R^2)} \ ; \\
   &\nabla_\mu k_\nu = \frac{i}{2R}g_{\mu\nu} - \frac{2i}{R}k_\mu k_\nu - \frac{2(1+R^2)}{R^2}k_{(\mu}r_{\nu)} \ ; \\
   &\nabla_\mu r_\nu = g_{\mu\nu} - 4k_\mu k_\nu + \frac{4i}{R}k_{(\mu}r_{\nu)} + \frac{1-R^2}{R^2}r_\mu r_\nu \ ; \\
   &\nabla_\mu k^\mu = \frac{i}{R} \ ; \quad k^\nu\nabla_\nu k_\mu = 0 \ ; \quad r^\nu\nabla_\nu k_\mu = -\frac{R^2 k_\mu}{1+R^2} \ ; \\
   &\Box \frac{1}{R} = -\frac{2}{R} \ ; \quad \Box k_\mu = -3k_\mu - \frac{i(1+R^2)}{R^3}r_\mu \ ; \\
   &\nabla^\rho k_\mu \nabla_\rho k_\nu = -\frac{1}{4R^2}g_{\mu\nu} + \frac{1+R^2}{R^2}k_\mu k_\nu - \frac{i(1+R^2)}{R^3}k_{(\mu}r_{\nu)} \ .
 \end{split} \label{eq:k_r_curved}
\end{align}
We also define a bivector in the $tr$ plane, as in \eqref{eq:flat_bivector} but with a curvature-corrected prefactor:
\begin{align}
 S^\perp_{\mu\nu} \equiv (1+R^2)t_{[\mu}r_{\nu]} = 2(1+R^2)k_{[\mu}r_{\nu]} \ , \label{eq:curved_bivector}
\end{align}
which satisfies identities very similar to \eqref{eq:flat_curls}:
\begin{align}
 k_{[\mu}\nabla_{\nu]} k_\rho = \frac{S^\perp_{\mu\nu} k_\rho}{2R^2} + \frac{i}{2R}k_{[\mu}g_{\nu]\rho} \ ; \quad \nabla_\mu S^\perp_{\nu\rho} = (1+R^2)t_{[\nu}g_{\rho]\mu} \ ,
\end{align}
and can be decomposed into left-handed and right-handed parts as:
\begin{align}
 S^{L/R}_{\mu\nu} \equiv \frac{1}{2}\left(S^\perp_{\mu\nu} \pm \frac{1}{2}\epsilon_{\mu\nu}{}^{\lambda\rho\sigma}x_\lambda S^\perp_{\rho\sigma}\right) \ .
\end{align}
With these building blocks in hand, it's easy to show that the solution to the field equation \eqref{eq:field_eqs_with_particle} in $EAdS_4$, as well as its Weyl curvature, take the same form as in the flat case:
\begin{align}
 h(x) &= -\frac{Q^{(0)}}{4\pi R} \quad (s=0) \ ; \label{eq:curved_solution_0} \\
 h_{\mu_1\dots\mu_s}(x) &= -\frac{Q^{(s)}}{2\pi R}\,k_{\mu_1}\!\dots k_{\mu_s} \quad (s\geq 1) \ ; \label{eq:curved_solution_s} \\
 \varphi_{\mu_1\nu_1\dots\mu_s\nu_s}(x) &= - \frac{(2s)!}{s!}\cdot\frac{Q^{(s)}}{4\pi R^{2s+1}}\left(S^L_{\mu_1\nu_1}\dots S^L_{\mu_s\nu_s} + S^R_{\mu_1\nu_1}\dots S^R_{\mu_s\nu_s}\right) \quad (s\geq 1) \ . \label{eq:curved_C}
\end{align}
To verify this, one must only repeat the calculations at $R\neq 0$. The analysis of the field equation at the $R=0$ singularity can be taken directly from the flat case, since the constant curvature of $EAdS_4$ becomes irrelevant at very short distances. For the $s\geq 1$ solution, we will again prefer the more symmetric gauge choice \eqref{eq:flat_solution_symm}:
\begin{align}
 h_{\mu_1\dots\mu_s}(x) = -\frac{Q^{(s)}}{2\pi R}\Re(k_{\mu_1}\!\dots k_{\mu_s}) \ . \label{eq:curved_solution_symm}
\end{align}
Finally, it will be useful to express $S^\perp_{\mu\nu}$ directly in terms of the worldline's boundary endpoints $\ell^\mu,\ell'^\mu$ and the bulk ``measurement point'' $x^\mu$. It turns out that $S^\perp_{\mu\nu}$ is just the projection of the bivector $\ell_{[\mu}\ell'_{\nu]}/(\ell\cdot\ell')$ into the tangent space at $x$, i.e. in perpendicular to $x^\mu$:
\begin{align}
 \begin{split}
   S_{\mu\nu} &\equiv \frac{\ell_{[\mu}\ell'_{\nu]}}{\ell\cdot\ell'} \ ; \\
   S^\perp_{\mu\nu} &= S_{\mu\nu} + x_\mu x^\rho S_{\rho\nu} + x_\nu x^\rho S_{\mu\rho} \ .
 \end{split} \label{eq:bivector_projection}
\end{align}

\section{HS interaction between two bulk particles} \label{sec:two_particles}

In this section, we study the action for two bulk particles with geodesic worldlines, interacting at leading order via HS fields. We will see that a particular pattern of charges leads to divergence cancellation and an especially simple result for the action. In section \ref{sec:boundary_bilocals}, we will identify this special pattern of charges as that of the Didenko-Vasiliev solution, and associate it with a bilocal operator on the boundary.

\subsection{General structure of the on-shell action}

In general, when we impose the field equations $G^{\mu_1\dots\mu_s} = T^{\mu_1\dots\mu_s}$, the two terms in the action \eqref{eq:S_with_current} become proportional to each other. The on-shell action then reads simply:
\begin{align}
 S = \frac{1}{2}\int d^4x\sqrt{g} \sum_{s=0}^\infty h_{\mu_1\dots\mu_s}T^{\mu_1\dots\mu_s} +  \text{boundary terms} \ , \label{eq:S_with_current_on_shell}
\end{align}
where we included a sum over spins. Specializing to a point-particle source, charged under the fields of different spins, this becomes:
\begin{align}
 S = \frac{1}{2}\int_\gamma d\tau \sum_{s=0}^\infty Q^{(s)} h_{\mu_1\dots\mu_s}u^{\mu_1}\!\dots u^{\mu_s} +  \text{boundary terms} \ . \label{eq:S_with_particle_on_shell_raw}
\end{align} 
We will consider here two particles, so really there should be a sum over two worldlines in \eqref{eq:S_with_particle_on_shell_raw}. However, we'll restrict our attention to the action due to the fields of \emph{one} particle acting on the \emph{other}. As usual, there will be an equal contribution from the second particle acting on the first; taking this into account cancels the factor of $\frac{1}{2}$ in \eqref{eq:S_with_particle_on_shell_raw}. The action of a particle's fields on itself, i.e. the particle's self-interaction, is typically UV-divergent. As we will see below, this isn't actually the case for an HS-charged particle with the Didenko-Vasiliev pattern of charges (though even then the action \emph{does} diverge if we restrict to even-spin charges only, i.e. if we average the charges between a particle and its antiparticle). At any rate, we'll treat the case of a particle acting on itself as just a special case of one particle acting on another. 

Two further subtelties should be addressed before the action \eqref{eq:S_with_particle_on_shell_raw} can be evaluated: the action's boundary terms, and its gauge-dependence. [UPDATE] The previous version of this paper confidently claimed a principled approach to these subtleties. In light of more recent work \cite{DidenkoVasilievAnalysis}, this is revealed to be mistaken. The most salient point from the previous version was that boundary terms should be absent, because the DV solution's field strength on the boundary is purely electric \cite{Neiman:2017mel}. However, this is true only away from the worldline endpoints, and we cannot offer a principled treatment that includes them. At the end of the day, in this paper, we simply compute the action in the gauge \eqref{eq:curved_solution_symm}, with no boundary terms, and observe in particular that it produces the holographically predicted result in eq. \eqref{eq:curved_result}. In our later work \cite{DidenkoVasilievAnalysis}, one can find an after-the-fact analytic proof of this result, which partially sheds light on the choice of gauge \eqref{eq:curved_solution_symm} and its relation with other gauges.

To summarize, we'll be evaluating an action of the form:
\begin{align}
 S = \frac{1}{2}\int_{\gamma_2} d\tau \sum_{s=0}^\infty Q_2^{(s)} u^{\mu_1}\!\dots u^{\mu_s} h_{\mu_1\dots\mu_s}(x_2;\gamma_1) \ , \label{eq:S_with_particle_on_shell}
\end{align}
with no boundary terms, where the integral is over the worldline of particle no. 2, and the field $h_{\mu_1\dots\mu_s}(x_2;\gamma_1)$ is the one generated by particle no. 1 at the location of particle no. 2. In particular, we will use $h_{\mu_1\dots\mu_s}$ in the gauge \eqref{eq:curved_solution_symm}, with the spin-0 case given separately by eq. \eqref{eq:curved_solution_0}. Plugging these in, the index contractions in \eqref{eq:S_with_particle_on_shell} reduce to powers of the scalar product $k_\mu u^\mu$, and the action takes the form:
\begin{align}
 \begin{split}
   S &= -\frac{1}{4\pi}\int_{\gamma_2} \frac{d\tau}{R}\left(\frac{1}{2}Q_1^{(0)}Q_2^{(0)} + \Re\sum_{s=1}^\infty Q_1^{(s)}Q_2^{(s)} (k_\mu u^\mu)^s \right) \\
      &= -\frac{1}{4\pi}\int_{\gamma_2} \frac{d\tau}{R}\left(\frac{1}{2}Q_1^{(0)}Q_2^{(0)} + \Re\sum_{s=1}^\infty \frac{Q_1^{(s)}Q_2^{(s)}}{2^s} \left(t_\mu u^\mu + \frac{ir_\mu u^\mu}{R}\right)^s \right) \ . 
 \end{split} \label{eq:S_with_particle_explicit} 
\end{align}
Here and below, the distance $R$, the radial vector $r^\mu$, the ``time'' vector $t^\mu$ and their null combination $k^\mu$ are defined with respect to worldline no. 1, and evaluated at the location of wordline no. 2; the 4-velocity $u^\mu$ is that of worldline no. 2. 

\subsection{The $\bbR^4$ case}

We begin in flat Euclidean spacetime. The two particles' worldlines are straight lines, at distance $b$ and angle $\theta$ (in the Lorentzian case, these would describe the particles' impact parameter and relative velocity). We can align the coordinate axes such that the two worldlines are situated at:
\begin{align}
 x_1^\mu(\tau) = (\tau,0,0,0) \ ; \quad x_2^\mu(\tau) = (\tau\cos\theta,\tau\sin\theta,b,0) \ .
\end{align}
The geometric ingredients of the action formula \eqref{eq:S_with_particle_explicit} then read:
\begin{align}
 \begin{split}
   t^\mu &= (1,0,0,0) \ ; \quad r^\mu = (0,\tau\sin\theta,b,0) \ ; \quad R = \sqrt{\tau^2\sin^2\theta + b^2} \ ; \\ 
   u^\mu &= (\cos\theta,\sin\theta,0,0) \ ; \quad t_\mu u^\mu = \cos\theta \ ; \quad r_\mu u^\mu = \tau\sin^2\theta \ ,
 \end{split}
\end{align}
so that the action takes the form:
\begin{align}
 S = -\frac{1}{4\pi}\int_{-\infty}^\infty \frac{d\tau}{\sqrt{\tau^2\sin^2\theta + b^2}}\left(\frac{1}{2}Q_1^{(0)}Q_2^{(0)} + \Re\sum_{s=1}^\infty \frac{Q_1^{(s)}Q_2^{(s)}}{2^s} \left(\cos\theta + \frac{i\tau\sin^2\theta}{\sqrt{\tau^2\sin^2\theta + b^2}}\right)^s \right) \ . 
 \label{eq:S_with_particle_flat_raw} 
\end{align}
This integral is scale-invariant. Upon switching to a dimensionless integration variable $\hat\tau \equiv \tau/b$, the $b$-dependence disappears:
\begin{align}
 S = -\frac{1}{4\pi}\int_{-\infty}^\infty \frac{d\hat\tau}{\sqrt{\hat\tau^2\sin^2\theta + 1}}\left(\frac{1}{2}Q_1^{(0)}Q_2^{(0)} + \Re\sum_{s=1}^\infty \frac{Q_1^{(s)}Q_2^{(s)}}{2^s} \left(\cos\theta + \frac{i\hat\tau\sin^2\theta}{\sqrt{\hat\tau^2\sin^2\theta + 1}}\right)^s \right) 
 \ . \label{eq:S_with_particle_flat} 
\end{align}
On the other hand, the integral is generally divergent as $\hat\tau \rightarrow \pm\infty$. For parallel or anti-parallel worldlines, i.e. $\theta=0,\pi$, we get a linear divergence:
\begin{align}
 S = -\frac{1}{4\pi b}\left(\frac{1}{2}Q_1^{(0)}Q_2^{(0)} + \Re\sum_{s=1}^\infty \left(\pm\frac{1}{2}\right)^s Q_1^{(s)}Q_2^{(s)} \right)\int_{-\infty}^\infty d\tau \ , \label{eq:S_parallel}
\end{align}
where the $\pm$ signs are for $\theta=0,\pi$ respectively, and for the moment we restored the dimensionful variables $\tau,b$. The general form of eq. \eqref{eq:S_parallel} is easy to understand: two particles at rest have some potential energy of interaction that scales as inverse distance $\sim 1/b$, and defines the action \emph{per worldline length}. Now, there exist particular combinations of charges $Q^{(s)}_{1,2}$ for which the coefficient in parentheses in \eqref{eq:S_parallel} vanishes; for these special combinations, the potential energy (and the resulting force) between two particles at rest is zero. The most famous example is a pair of extremal charged black holes in Einstein-Maxwell theory, or BPS particles in supergravity. For two such objects at rest, the electric repulsion precisely cancels the gravitational attraction. In our notation, this corresponds to the case $\theta = 0$, with particles charged only under the $s=1$ gauge field (electric charge) and the $s=2$ field (gravitational mass), with the charges related as $Q^{(1)} = \pm\frac{i}{\sqrt{2}}Q^{(2)}$. Here, the imaginary electric charge is a standard consequence of working in Euclidean signature. At $\theta = \pi$, these same charges yield contributions that add up rather than cancel; in Lorentzian signature, this corresponds to a particle and \emph{anti}particle at rest, with both electric and gravitational forces attractive.

The integral \eqref{eq:S_with_particle_flat} diverges also for general angles $0<\theta<\pi$, but logarithmically rather than linearly. Specifically, at both ends $\tau\rightarrow \pm\infty$ of the worldline, the integral takes the form:
\begin{align}
 S_{\text{log-divergent}} = -\frac{1}{4\pi\sin\theta}\left(\frac{1}{2}Q_1^{(0)}Q_2^{(0)} + \sum_{s=1}^\infty \frac{Q_1^{(s)}Q_2^{(s)}}{2^s}\cos(s\theta) \right) \int \frac{d\tau}{|\tau|} \ . \label{eq:S_log_div}
\end{align}
Comparing with \eqref{eq:S_parallel}, we can see that the introduction of an angle (i.e. a relative velocity) brings out the different tensor structures of the different-spin interactions, in the form of the angle-dependent $\cos(s\theta)$ factors. In particular, the $s=1$ and $s=2$ contributions to the logarithmic divergence have different $\theta$ dependence, and therefore can no longer cancel each other; in particular, for the BPS charge assignment $Q^{(1)} = \frac{i}{\sqrt{2}}Q^{(2)}$, we get a cancellation only at $\theta = 2\pi/3$.

On the other hand, if we let go of the restriction to spins $s=1,2$, we can obtain a cancellation of the divergences at \emph{almost all angles}. Suppose, as in the BPS case, that both particles have the same proportionality pattern between the charges of different spins, i.e.:
\begin{align}
 \frac{Q_1^{(s)}}{Q_1^{(0)}} =  \frac{Q_2^{(s)}}{Q_2^{(0)}} \equiv q_s \ ,
\end{align}
The logarithmic divergence \eqref{eq:S_log_div} of the action then reads:
\begin{align}
 S_{\text{log-divergent}} = -\frac{Q_1^{(0)}Q_2^{(0)}}{4\pi\sin\theta}\left(\frac{1}{2} + \sum_{s=1}^\infty \frac{q_s^2}{2^s}\cos(s\theta) \right) \int \frac{d\tau}{|\tau|} \ . \label{eq:S_log_div_identical}
\end{align}
We see that the squared charges of the different spins act as Fourier coefficients for the $\theta$-dependence of the logarithmic divergence. This means that we cannot quite cancel the divergence for all $\theta$, but we can increase the domain of cancellation from $\theta=0$ all the way to $0\leq\theta < \pi$, by making the expression in parentheses in \eqref{eq:S_log_div_identical} proportional to $\delta(\theta - \pi)$:
\begin{align}
 \frac{1}{2} + \sum_{s=1}^\infty \frac{q_s^2}{2^s}\cos(s\theta) \ \sim \ \delta(\theta - \pi) \ .
\end{align} 
This is accomplished by choosing $q_s^2 = (-2)^s$, i.e.:
\begin{align}
 Q^{(s)} = \pm (i\sqrt{2})^s Q^{(0)} \ , \label{eq:charge_pattern}
\end{align}
which is consistent with the BPS assignment for $s=1,2$, but extends it to all spins. As we will see in section \ref{sec:boundary_bilocals}, the pattern of charges \eqref{eq:charge_pattern} coincides with the one for the Didenko-Vasiliev black hole. We therefore refer to it as the DV pattern. Plugging it back into the full action formula \eqref{eq:S_with_particle_flat}, we find that the sum over spins becomes a geometric series. Summing the series, we arrive at an integral that can be performed analytically:
\begin{align}
 \begin{split}
  S &= -\frac{Q_1^{(0)}Q_2^{(0)}(1 - \cos\theta)}{8\pi(1 + \cos\theta)}\int_{-\infty}^\infty \frac{d\hat\tau}{\big(2\hat\tau^2(1 - \cos\theta) + 1 \big)\sqrt{\hat\tau^2\sin^2\theta + 1}} \\
     &=-\frac{Q_1^{(0)}Q_2^{(0)}}{8\pi(1+\cos\theta)}\left.\arctan\frac{(1-\cos\theta)\hat\tau}{\sqrt{\hat\tau^2\sin^2\theta + 1}}\right|_{\hat\tau = -\infty}^\infty \\
     &= -\frac{Q_1^{(0)}Q_2^{(0)}\theta}{8\pi(1+\cos\theta)} \ . 
 \end{split} \label{eq:flat_result}
\end{align}
As anticipated above, we see that the action vanishes for $\theta = 0$, diverges for $\theta = \pi$, and is finite for all intermediate values $0<\theta<\pi$. To recapitulate, this finiteness is due to cancellations of the divergences \eqref{eq:S_parallel}-\eqref{eq:S_log_div}, in a higher-spin-enhanced version of the cancellation for BPS particles in supergravity, which takes place only at $\theta = 0$ and $\theta=2\pi/3$.

We emphasize again that the dependence on the impact parameter $b$ dropped out of the action \eqref{eq:flat_result}, due to scale invariance. This scale invariance enables us to take two equivalent viewpoints on the cancelled divergences. Our calculation above made them appear as IR divergences: at fixed impact parameter $b$, the action diverges as we integrate over distant portions of the worldline. However, these same divergences can be viewed as UV ones: if we cut off the integration at some fixed distance along the worldline, then the action becomes finite at fixed $b$, but diverges for $b\rightarrow 0$, i.e. when the two particles collide.

\subsection{The $EAdS_4$ case}

\begin{figure}%
	\centering%
	\includegraphics[scale=0.6]{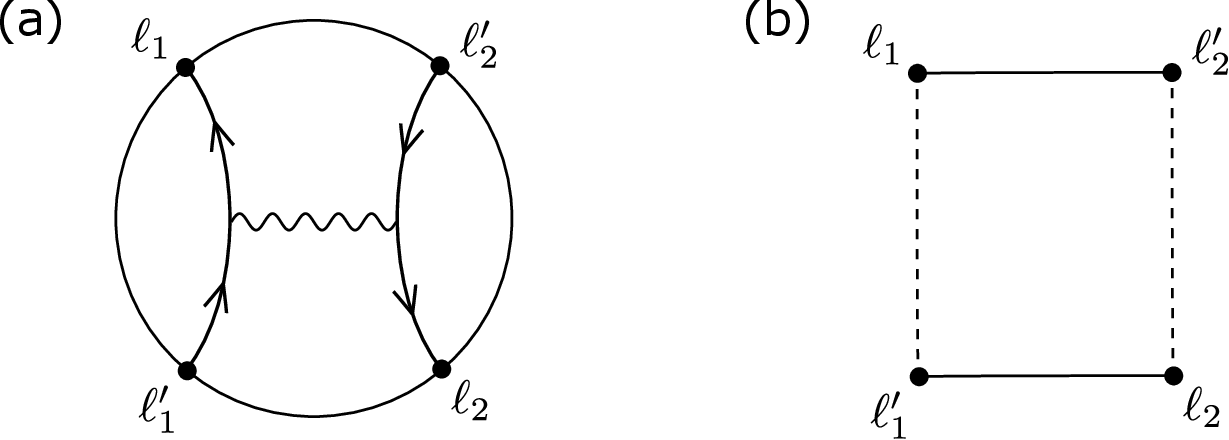} \\
	\caption{(a) Two HS-charged particles, traveling along geodesics in $EAdS_4$, are interacting via their HS gauge fields. (b) A Feynman diagram in the boundary vector model, computing the correlator of two bilocal operators $\bar\phi_I(\ell'_1)\phi^I(\ell_1)$ and $\bar\phi_I(\ell'_2)\phi^I(\ell_2)$; the solid lines are propagators, while the dashed lines are just contractions of the color indices $I$. The bulk picture in (a), with each of the two particles assigned the DV pattern of charges, describes an HS multiplet of Witten diagrams that compute the same correlator. The boundary diagram in (b) is \emph{almost} one of the Feynman diagrams for the 4-point function of scalar operators $j^{(0)}(\ell) = \bar\phi_I(\ell)\phi^I(\ell)$, but with two propagators missing.}
	\label{fig:4pt} 
\end{figure}

We now turn to the case of two particles interacting via HS fields in $EAdS_4$ (see figure \ref{fig:4pt}a). Our first task is again to parameterize the particles' geodesic worldlines. The relative position of the worldlines is again characterized by two numbers: an impact parameter, which we will now denote by $\chi$, and a relative angle $\theta$. $\chi$ is defined as the length of the shortest interval in $EAdS_4$ connecting the two worldlines; it is a hyperbolic angle in the $\bbR^{1,4}$ embedding space. $\theta$ is defined as the angle between the worldlines' 4-velocities, evaluated at the ends of this shortest interval (the angle can be defined equivalently either in embedding space, or intrinsically in $EAdS_4$ using parallel transport along the interval).

In a suitably chosen Lorentz frame in the $\bbR^{1,4}$ embedding space, we can fix the positions and 4-velocities of the two worldlines at their closest points as:
\begin{align}
 \begin{split}
   x_1^\mu(0) &= (1,0,0,0,0) \ ; \quad u_1^\mu(0) = (0,1,0,0,0) \ ; \\
   x_2^\mu(0) &= (\cosh\chi,0,\sinh\chi,0,0) \ ; \quad u_2^\mu(0) = (0,\cos\theta,0,\sin\theta,0) \ .
 \end{split} 
\end{align}
From these, we can construct the worldlines themselves as:
\begin{align}
 \begin{split}
   x_1^\mu(\tau) &= x_1^\mu(0)\cosh\tau + u_1^\mu(0)\sinh\tau \\
     &= (\cosh\tau,\sinh\tau,0,0,0) \ ;
 \end{split} \label{eq:x_1} \\
 \begin{split}
   x_2^\mu(\tau) &= x_2^\mu(0)\cosh\tau + u_2^\mu(0)\sinh\tau \\
     &= (\cosh\chi\cosh\tau,\cos\theta\sinh\tau,\sinh\chi\cosh\tau,\sin\theta\sinh\tau,0) \ .
 \end{split} \label{eq:x_2}
\end{align}
In particular, the 4-velocity along the 2nd worldline reads:
\begin{align}
 u^\mu(\tau) \equiv u_2^\mu(\tau) = (\cosh\chi\sinh\tau,\cos\theta\cosh\tau,\sinh\chi\sinh\tau,\sin\theta\cosh\tau,0) \ .
\end{align}
Taking the limits $\tau\rightarrow \pm\infty$ in \eqref{eq:x_1}-\eqref{eq:x_2} and extracting coefficients of $\frac{1}{2}e^{|\tau|}$, we identify the boundary endpoints of the two worldlines as:
\begin{align}
 \ell_1^\mu &= (1,1,0,0,0) \ ; \quad \ell'^\mu_1 = (1,-1,0,0,0) \ ; \label{eq:ell_1} \\
 \ell_2^\mu &= (\cosh\chi,\cos\theta,\sinh\chi,\sin\theta,0) \ ; \quad \ell'^\mu_2 = (\cosh\chi,-\cos\theta,\sinh\chi,-\sin\theta,0) \ .
\end{align}
Our two parameters $\chi$ and $\theta$ are just a particular encoding of the two independent cross-ratios of the four boundary points $\ell_1,\ell'_1,\ell_2,\ell'_2$. Explicitly:
\begin{align}
 \sqrt{\frac{(\ell_1\cdot\ell'_2)(\ell_2\cdot\ell'_1)}{(\ell_1\cdot\ell'_1)(\ell_2\cdot\ell'_2)}} = \frac{1}{2}(\cosh\chi + \cos\theta) \ ; \quad \sqrt{\frac{(\ell_1\cdot\ell_2)(\ell'_1\cdot\ell'_2)}{(\ell_1\cdot\ell'_1)(\ell_2\cdot\ell'_2)}} = \frac{1}{2}(\cosh\chi - \cos\theta)  \ .
\end{align}
Plugging the 2nd worldline's position \eqref{eq:x_2} and the 1st worldline's endpoints \eqref{eq:ell_1} into eqs. \eqref{eq:R}-\eqref{eq:r_t}, we obtain the ingredients of the action integral \eqref{eq:S_with_particle_explicit} as:
\begin{align}
 \begin{split}
   R &= \sqrt{(\cosh^2\chi - \cos^2\theta)\sinh^2\tau + \sinh^2\chi} \ ; \\
   r_\mu u^\mu &= \frac{(\cosh^2\chi - \cos^2\theta)\cosh\tau\sinh\tau}{(\cosh^2\chi - \cos^2\theta)\sinh^2\tau + \cosh^2\chi} \ ; \\ 
   t_\mu u^\mu &= \frac{\cosh\chi\cos\theta}{(\cosh^2\chi - \cos^2\theta)\sinh^2\tau + \cosh^2\chi} \ .
 \end{split}
\end{align}
We can simplify these expressions somewhat by switching variables from $\tau$ (which ranges from $-\infty$ to $\infty$) to $R$ (which ranges twice from $\sinh\chi$ to $\infty$). Plugging everything into the action integral \eqref{eq:S_with_particle_explicit}, we get:
\begin{align}
 \begin{split}  
  S ={}& {-\frac{1}{2\pi}}\int_{\sinh\chi}^\infty \frac{dR}{\sqrt{(R^2 - \sinh^2\chi)(R^2 + \sin^2\theta)}} \Bigg(\frac{1}{2}Q_1^{(0)}Q_2^{(0)} + {} \\
   &\ + \Re\sum_{s=1}^\infty \frac{Q_1^{(s)}Q_2^{(s)}}{2^s(R^2+1)^s} \left(\cosh\chi\cos\theta + \frac{i}{R}\sqrt{(R^2 - \sinh^2\chi)(R^2 + \sin^2\theta)} \right)^s\,  \Bigg) \ ,
 \end{split} \label{eq:S_curved} 
\end{align}
The integral \eqref{eq:S_curved} does not have the scale-invariance of its $\bbR^4$ counterpart, due to the $EAdS_4$ curvature radius (here, set to 1). Due to the negative curvature, geodesics in $EAdS_4$ recede from each other at large distances much faster than in $\bbR^4$. As a result, unlike its flat counterpart, the integral \eqref{eq:S_curved} is IR-finite for any assignment of charges $Q_{1,2}^{(s)}$: at $R\rightarrow\infty$, the spin-$s$ piece of the integral goes as $\int dR/R^{s+2}$. On the other hand, there are still UV divergences in the limit $\chi\rightarrow 0$, i.e. as the two particles collide. Since the spacetime curvature becomes negligible at short distances, these UV divergences are the same as the ones we studied in the $\bbR^4$ case. They can thus be cancelled, for all values of $\theta$ except $\theta = \pi$, by assigning the DV pattern of charges \eqref{eq:charge_pattern} to both particles. With this assignment, the sum over spins in \eqref{eq:S_curved} again becomes a geometric sum. Performing it, we bring the action integral into the form:
\begin{align}
 \begin{split}   
   S ={}& \frac{Q_1^{(0)}Q_2^{(0)}}{2\pi}\int_{\sinh\chi}^\infty \frac{dR}{\sqrt{(R^2 - \sinh^2\chi)(R^2 + \sin^2\theta)}} \Bigg(\frac{1}{2} -{} \\
   &\, - (R^2+1)\Re \left(R^2 + 1 + \cosh\chi\cos\theta + \frac{i}{R}\sqrt{(R^2 - \sinh^2\chi)(R^2 + \sin^2\theta)} \right)^{-1}\,  \Bigg) \ .
 \end{split} \label{eq:S_curved_DV} 
\end{align}
We've been unable to significantly simplify this integral, or to evaluate it analytically. One might have hoped to at least use the flat result \eqref{eq:flat_result} in the limit $\chi\rightarrow 0$ when the particles come very close together, and extrapolate from there. However, even in this limit, the $EAdS_4$ result \emph{isn't} captured by the $\bbR^4$ one, precisely due to the cancellation of UV divergences. On the other hand, as we'll discuss below, holography predicts a very simple answer for the integral \eqref{eq:S_curved_DV}:
\begin{align}
 S = -\frac{Q_1^{(0)}Q_2^{(0)}}{8(\cosh\chi + \cos\theta)} = -\frac{Q_1^{(0)}Q_2^{(0)}}{16}\sqrt{\frac{(\ell_1\cdot\ell'_1)(\ell_2\cdot\ell'_2)}{(\ell_1\cdot\ell'_2)(\ell_2\cdot\ell'_1)}} \ . \label{eq:curved_result}
\end{align}
We've verified the agreement between \eqref{eq:S_curved_DV} and \eqref{eq:curved_result} by numerical integration in Mathematica, for various values of the parameters $\chi,\theta$. The formula \eqref{eq:curved_result} for the leading-order interaction of two HS-charged particles in $EAdS_4$ with the DV pattern of charges is the main technical result of our paper. In the next sections, we will place our analysis of HS-charged particles in a broader context, by connecting it to the Didenko-Vasiliev black hole solution, as well as to higher-spin holography.

\section{Twistors, HS algebra and boundary correlators} \label{sec:twistors}

From here on, we focus on the case of $EAdS_4$ spacetime. In this section, we introduce the tools necessary to connect our results above with the Didenko-Vasiliev solution and with AdS/CFT. In essence, we need to switch from the Fronsdal's ``metric-like'' formulation of HS fields to Vasiliev's language of twistors, HS algebra and master fields. More specifically, we will introduce these in a slightly non-standard approach, developed by us in \cite{Neiman:2013hca,Neiman:2015wma,Neiman:2017mel,David:2020ptn}. The idea is to work in the embedding-space picture, and to introduce twistors in a way that is closer to Penrose's original sense of the word \cite{Penrose:1986ca,Ward:1990vs} -- as spinors of the spacetime symmetry group $SO(1,4)$, that exist without tethering to any particular spacetime point. 

In section \ref{sec:twistors:intro}, we review twistor space, HS algebra, bulk master fields and the Penrose transform. In section \ref{sec:twistors:correlators}, we review the free vector model that lives on the conformal boundary of $EAdS_4$, and the HS-algebraic generating function for its correlators. In section \ref{sec:twistors:normalization}, we use the boundary 2-point functions to fix the relative normalizations between the Fronsdal and twistor languages. The content of sections \ref{sec:twistors:intro}-\ref{sec:twistors:correlators} is a telegraphic summary of constructions detailed at length in \cite{Neiman:2017mel,David:2020ptn}; the calculation in section \ref{sec:twistors:normalization} is new.

\subsection{Twistor space, HS algebra and the Penrose transform} \label{sec:twistors:intro}

For the purposes of this paper, twistor space is the space of (4-component, Dirac) spinors of the $EAdS_4$ isometry group $SO(1,4)$. In other words, twistors are the spinors of the $\bbR^{1,4}$ embedding space. We use Latin indices $(a,b,\dots)$ to denote twistors. Twistor space has a symplectic metric $I_{ab}$ with inverse $I^{ab}I_{ac} = \delta^b_c$, which we use to raise and lower indices as $U_a = I_{ab}U^b$, $U^a = U_b I^{ba}$. It is often convenient to use index-free notation, in which bottom-to-top index contraction is implied, e.g. $UV \equiv U_a V^a$. The translation between twistors and tensors is performed by the Dirac gamma matrices $(\gamma_\mu)^a{}_b$, which satisfy the Clifford algebra $\gamma_{(\mu}\gamma_{\nu)} = -\eta_{\mu\nu}$ (these are just the familiar gamma matrices from $\bbR^{1,3}$, with the addition of $\gamma_5$). It is also useful to define the antisymmetric combinations $\gamma_{\mu\nu}\equiv \gamma_{[\mu}\gamma_{\nu]}$, which generate the $SO(1,4)$ spacetime symmetry within Clifford algebra. The matrices $\gamma_\mu^{ab}$ are antisymmetric and traceless in their twistor indices, while the $\gamma_{\mu\nu}^{ab}$ are symmetric. We define the following dictionaries between objects with tensor and twistor indices:
\begin{align}
 \xi^{ab} = \gamma_\mu^{ab}\xi^\mu \ ; \quad \xi^\mu = -\frac{1}{4}\gamma^\mu_{ab}\xi^{ab} \ ; \quad 
 \zeta^{ab} = \frac{1}{2}\gamma_{\mu\nu}^{ab}\zeta^{\mu\nu} \ ; \quad \zeta^{\mu\nu} = \frac{1}{4}\gamma^{\mu\nu}_{ab}\zeta^{ab} \ .
\end{align}
We now define HS algebra in complete analogy to Clifford algebra. Instead of a vector of quantities $\gamma^\mu$ whose anticommutator is given by the spacetime metric $\eta_{\mu\nu}$, we define a \emph{twistor} variable $Y^a$ whose \emph{commutator} is given by the twistor metric $I_{ab}$. We denote this non-commutative product with a star $\star$, and realize it as:
\begin{align}
 Y^a\star Y^b = Y^a Y^b + iI^{ab} \ . \label{eq:star}
\end{align}
The product \eqref{eq:star} is easily extended to an associative product on twistor polynomials $f(Y)$. A further generalization to arbitrary functions $f(Y)$ is possible, and is formally given by an integral formula:
\begin{align}
 f(Y)\star g(Y) = \int d^4U d^4V f(Y+U)\, g(Y+V)\, e^{-iUV} \ , \label{eq:star_int}
\end{align}
where the twistor integration measure is defined as:
\begin{align}
 d^4U \equiv \frac{\epsilon_{abcd}}{4!(2\pi)^2}\,dU^a dU^b dU^c dU^d \ ; \quad \epsilon_{abcd} \equiv 3I_{[ab}I_{cd]} \ .
\end{align}
The algebra defined by the product \eqref{eq:star}-\eqref{eq:star_int} (restricted to even functions, i.e. to integer spins) is known as HS algebra, and defines the infinite-dimensional symmetry group of HS theory. Strictly speaking, to define an HS algebra properly, one must restrict to an appropriate class of functions (or distributions) $f(Y)$, worry about boundary conditions etc.; see e.g. \cite{DeFilippi:2019jqq}. These issues won't concern us here.

HS algebra contains the spacetime symmetry $SO(1,4)$ as a finite-dimensional subalgebra. The latter is generated, just as in Clifford algebra, by the quadratic elements $Y_a Y_b$. HS algebra admits a trace operation, defined simply by:
\begin{align}
 \tr_\star f(Y) = f(0) \ .
\end{align}
So far, we made no reference to any spacetime points. Choices of spacetime points, either in the bulk of $EAdS_4$ or on its boundary, induce decompositions of twistor space, and thus of HS algebra. A choice of bulk point $x^\mu$ decomposes twistor space into right-handed and left-handed Weyl spinor spaces at $x$, via the projectors:
\begin{align}
 P^{ab}(\pm x) = \frac{1}{2}\left(I^{ab} \pm x^\mu\gamma^{ab}_\mu \right) \ , \label{eq:P_x}
\end{align}
or, in index-free notation, simply $P(\pm x) = \frac{1}{2}(1\pm x)$. We use the same notation $P(\pm x)$ to denote the two spinor spaces themselves. The two spinor spaces at $x$ are orthogonal to each other under the twistor metric, which simply decomposes as $I_{ab} = P_{ab}(x) + P_{ab}(-x)$. We denote the right-handed and left-handed Weyl-spinor pieces of a twistor $U^a$ at $x$ as $u_{(\pm x)} \equiv P(\pm x)U$. We define a measure on each spinor space, and a corresponding delta function, as:
\begin{align}
 d^2u_{(\pm x)}  = \frac{P_{ab}(\pm x)}{2(2\pi)}\,du_{(\pm x)}^a du_{(\pm x)}^b \ ; \quad \delta_{\pm x}(Y) = \int_{P(\pm x)} d^2u\,e^{iuY} \ .
\end{align}

A key concept in twistor theory is the \emph{Penrose transform}, which maps between twistor functions and solutions to free massless field equations in 4d. It was noticed in \cite{Neiman:2017mel} that the Penrose transform has a very elegant expression in HS theory. Specifically, we can map between an even twistor function $f(Y)$ and a master field $C(x;Y)$ in $EAdS_4$ that contains an HS multiplet of free massless fields, via a simple star product:
\begin{align}
 C(x;Y) = if(Y)\star \delta_x(Y) \ . \label{eq:Penrose}
\end{align}
When written out explicitly, the star product in \eqref{eq:Penrose} is a Fourier transform of the right-handed spinor $y_{(x)}$ (the Penrose transform is famously chiral; of course, a left-handed transform can also be defined). The object $C(x;Y) = C(x;y_{(x)}+y_{(-x)})$ is the usual linearized zero-form master field from the HS literature, up to some nuances of the formalism, which we return to in section \ref{sec:boundary_bilocals:comparing_formalisms}. In particular, it acts as a generating function for the Weyl-like field strengths \eqref{eq:C} of HS fields of all spins, together with their derivatives. The field strengths (as opposed to their derivatives) are contained in the purely chiral parts $C(x;y_{(x)})$ and $C(x;y_{(-x)})$ of the master field, as:
\begin{align}
 \begin{split}
   C_{\mu_1\nu_1\dots\mu_s\nu_s}(x) &= \frac{1}{4^s}\gamma_{\mu_1\nu_1}^{a_1 a_2}\dots \gamma_{\mu_s\nu_s}^{a_{2s-1} a_{2s}}\left( C^R_{a_1 b_1\dots a_s b_s}(x) + C^L_{a_1 b_1\dots a_s b_s}(x) \right) \ ; \\
   C^R_{a_1 b_1\dots a_s b_s}(x) &= \left.\frac{\del^{2s} C(x;y_{(x)})}{\del y_{(x)}^{a_1}\dots\del y_{(x)}^{a_{2s}}}\right|_{y_{(x)}=0} \ ; \\
   C^L_{a_1 b_1\dots a_s b_s}(x) &= \left.\frac{\del^{2s} C(x;y_{(-x)})}{\del y_{(-x)}^{a_1}\dots\del y_{(-x)}^{a_{2s}}}\right|_{y_{(-x)}=0} \ ,
 \end{split} \label{eq:Penrose_s}
\end{align}
where the spinors $C^R_{a_1 b_1\dots a_s b_s}(x)$ and $C^L_{a_1 b_1\dots a_s b_s}(x)$ encode the right-handed and left-handed parts of the field strength at $x$, respectively. The spin-0 field is simply given by:
\begin{align}
 C(x) = C(x;0) \ . \label{eq:Penrose_0}
\end{align}
The rest of the master field's Taylor expansion, with non-zero powers of both $y_{(x)}$ and $y_{(-x)}$, encodes derivatives of the field strengths \eqref{eq:Penrose_s}:
\begin{align}
 \begin{split}
   \left.\frac{\del^{2(s+k)} C(x;Y)}{\del y_{(x)}^{a_1}\dots\del y_{(x)}^{a_{2s+k}}\del y^{(-x)}_{b_1}\dots\del y^{(-x)}_{b_k}}\right|_{Y=0} 
     &= i^k\, \hat\nabla^{(b_1}{}_{(a_1}\dots\hat\nabla^{b_k)}{}_{a_k} C^R_{a_{k+1}\dots a_{2s+k})}(x) \ ; \\
   \left.\frac{\del^{2(s+k)} C(x;Y)}{\del y_{(x)}^{a_1}\dots\del y_{(x)}^{a_k}\del y^{(-x)}_{b_1}\dots\del y^{(-x)}_{b_{2s+k}}}\right|_{Y=0} 
     &= i^k\, \hat\nabla^{(b_1}{}_{(a_1}\dots\hat\nabla^{b_k}{}_{a_k)} C_L^{\,b_{k+1}\dots b_{2s+k})}(x) \ ,
 \end{split} \label{eq:unfolding}
\end{align}
where the covariant derivative with spinor indices is defined as:
\begin{align}
 \hat\nabla_{ab} \equiv P^c{}_a(-x) P^d{}_b(x) \gamma^\mu_{cd} \nabla_\mu \ ,
\end{align}
forcing the left-handed spinor index into the first position, and the right-handed one into the second position.

The transform \eqref{eq:Penrose} automatically enforces the derivative relations \eqref{eq:unfolding}, and also ensures that the fields \eqref{eq:Penrose_s}-\eqref{eq:Penrose_0} satisfy the appropriate field equations \eqref{eq:C_field_eqs} in $EAdS_4$. As usual in HS theory, all these equations can be encoded as a single unfolded equation on $C(x;Y)$ itself -- see section 5C of \cite{Neiman:2017mel}, or section 3B of \cite{Neiman:2015wma}; for our formalism's version of the full non-linear Vasiliev equations, see section 4 of \cite{Neiman:2015wma}. 

Note that our notation $C_{\mu_1\nu_1\dots\mu_s\nu_s}(x)$ for the field strengths in \eqref{eq:Penrose_s}-\eqref{eq:Penrose_0} is different from the one we used so far, i.e. $\varphi_{\mu_1\nu_1\dots\mu_s\nu_s}(x)$. We reserve the latter notation for field strengths derived, via \eqref{eq:C}, from potentials with a \emph{canonically normalized kinetic term}, as in \eqref{eq:S}. So far, we haven't given the twistor function $f(Y)$ and the Penrose-transformed fields \eqref{eq:Penrose}-\eqref{eq:Penrose_0} a meaningful normalization. In section \ref{sec:twistors:correlators}, we will equip them with one, by tying them to holographic correlators. In section \ref{sec:twistors:normalization}, we will work out the proportionality coefficients between these fields and the canonically normalized ones from \eqref{eq:C}.

\subsection{Boundary correlators from twistor functions} \label{sec:twistors:correlators}

Here, we begin to turn our attention to the holographic CFT dual of HS gravity, which lives on the boundary of $EAdS_4$. In the simplest case that we're considering, this CFT is a free vector model of $N$ complex massless scalar fields $\phi^I$, subject to $U(N)$ symmetry. Its single-trace primary operators are a tower of conserved currents, one for each spin $s$:
\begin{align}
 j^{(s)}_{k_1\dots k_s} = \frac{1}{i^s}\,\bar\phi_I \left(\sum_{m=0}^s (-1)^m \binom{2s}{2m} \overset{\leftarrow}{\del}_{(k_1}\dots\overset{\leftarrow}{\del}_{k_m} \overset{\rightarrow}{\del}_{k_{m+1}}\dots\overset{\rightarrow}{\del}_{k_s)} - \text{traces}\right) \phi^I \ ,
 \label{eq:j}
\end{align}
whose bulk duals are the HS gauge fields. One can uplift the 3d boundary indices into 5d indices in the $\bbR^{1,4}$ embedding space. Also, it is convenient to package the tensor components of \eqref{eq:j} at a boundary point $\ell$ by contracting with a null polarization vector $\lambda^\mu$, like the one we introduced in section \ref{sec:Fronsdal:propagators} (satisfying $\lambda\cdot\ell = \lambda\cdot\lambda = 0$):
\begin{align}
 \begin{split}
   j^{(s)}(\ell,\lambda) &= \lambda^{\mu_1}\dots\lambda^{\mu_s} j_{\mu_1\dots\mu_s}(\ell) \\
      &= \frac{\lambda^{\mu_1}\dots\lambda^{\mu_s}}{i^s}\,
        \bar\phi_I(\ell) \sum_{m=0}^s (-1)^m \binom{2s}{2m} \overset{\leftarrow}{\del}_{(\mu_1}\dots\overset{\leftarrow}{\del}_{\mu_m} \overset{\rightarrow}{\del}_{\mu_{m+1}}\dots\overset{\rightarrow}{\del}_{\mu_s)}\,\phi^I(\ell) \ .
 \end{split} \label{eq:contracted_j}
\end{align}
As discussed in \cite{Neiman:2017mel,David:2020ptn}, there exists a ``holographic dual of the Penrose transform'': a dictionary that encodes single-trace operator insertions in the CFT as twistor functions. In turn, these twistor functions correspond via the (ordinary, bulk) Penrose transform to linearized bulk fields with the appropriate boundary data. In terms of these twistor functions $f(Y)$, the generating function for the CFT correlators is given by the HS-algebraic expression:
\begin{align}
 Z[f(Y)] = \exp\left(\frac{N}{4}\sum_{n=1}^\infty \frac{(-1)^{n+1}}{n}\,\tr_\star\Big(\underbrace{f(Y)\star\ldots\star f(Y)}_{n\text{ factors}}\Big) \right) \ . \label{eq:Z}
\end{align}
This partition function defines the on-shell bulk action of HS gravity (at least in the classical limit, i.e. at large $N$). In particular, it lends meaning to the normalization of the bulk fields \eqref{eq:Penrose}-\eqref{eq:Penrose_0} produced from $f(Y)$ via the Penrose transform.

To make this more explicit, let us write down the twistor function that corresponds to the boundary current \eqref{eq:contracted_j}. First, we must briefly discuss the structure imposed on twistor space by a choice of boundary point $\ell$. At a bulk point, we saw that twistor space decomposes into the two chiral subspaces \eqref{eq:P_x}. At a boundary point $\ell$, only a single 2d subspace is singled out -- the subspace $P(\ell)$ spanned by $\ell^{ab} = \ell^\mu\gamma_\mu^{ab}$. This ends up serving as the space of 2-component \emph{cospinors} on the 3d boundary. Though $P(\ell)$ is totally null under the twistor metric $I_{ab}$, one can equip it with a symplectic metric, or equivalently a measure $d^2u_{(\ell)}$, by using $\ell^{ab}$ itself:
\begin{align}
 \frac{du_{(\ell)}^a du_{(\ell)}^b}{2\pi} \equiv \frac{1}{2}\ell^{ab}\, d^2u_{(\ell)} \ .
\end{align}
This metric scales under rescalings of $\ell^\mu$, as is appropriate for a metric on the conformal boundary. We can use it to define a delta function with support on $P(\ell)$:
\begin{align}
 \delta_\ell(Y) = \int d^2u_{(\ell)}\, e^{iu_{(\ell)}Y} \ .
\end{align}
The twistor function corresponding to the boundary current \eqref{eq:contracted_j} is constructed from this delta function as \cite{David:2020ptn}:
\begin{align}
 \kappa^{(s)}(\ell,\lambda;Y) = \frac{iM^{a_1}\dots M^{a_{2s}}}{8\pi} \left(Y_{a_1}\dots Y_{a_{2s}} + (-1)^s\frac{\del^{2s}}{\del Y^{a_1}\dots\del Y^{a_{2s}}}\right) \delta_\ell(Y) \ , \label{eq:kappa_s}
\end{align}
where $M^a$ is a polarization spinor, defined as an appropriate square root of the bivector $M^{\mu\nu} \equiv 2\ell^{[\mu}\lambda^{\nu]}$:
\begin{align}
 \gamma_{\mu\nu}^{ab}\ell^\mu\lambda^\nu = \frac{1}{2}\gamma_{\mu\nu}^{ab}M^{\mu\nu} = (\ell M)^a(\ell M^b) \ .
\end{align}
The Penrose transform \eqref{eq:Penrose} of the twistor function \eqref{eq:kappa_s} reads \cite{David:2020ptn}:
\begin{align}
 C(x;Y) = \frac{1}{4\pi}\cdot\frac{(M\ell P_{-x} Y)^{2s} + (M\ell P_x Y)^{2s}}{(\ell\cdot x)^{2s+1}}\exp\frac{iY\ell xY}{2(\ell\cdot x)} \ .
\end{align} 
Here, the field strength at $x$ is contained in the exponent's prefector, while the exponent itself carries the tower of derivatives. Explicitly, the field strength, extracted via eqs. \eqref{eq:Penrose_s}-\eqref{eq:Penrose_0}, reads:
\begin{align}
 C_{\mu_1\nu_1\dots\mu_s\nu_s}(x) = \frac{(2s)!}{4\pi}\cdot\frac{M^L_{\mu_1\nu_1}\dots M^L_{\mu_s\nu_s} + M^R_{\mu_1\nu_1}\dots M^R_{\mu_s\nu_s}}{(\ell\cdot x)^{2s+1}} \ , \label{eq:boundary_bulk_C}
\end{align}
where $M^{L/R}_{\mu_1\nu_1}$ are the projections of $M_{\mu\nu}$ onto the left-handed and right-handed bivector spaces at $x$, as in \eqref{eq:M_chiral}. The field strength \eqref{eq:boundary_bulk_C} clearly coincides, up to numerical factors, with the boundary-to-bulk propagator \eqref{eq:boundary_bulk_field_strength_0}-\eqref{eq:boundary_bulk_field_strength_s}. The $s=0$ case is included in \eqref{eq:kappa_s} and \eqref{eq:boundary_bulk_C}, as:
\begin{align}
 \kappa^{(0)}(\ell;Y) &= \frac{i}{4\pi}\delta_\ell(Y) \ ; \label{eq:kappa_0} \\
 C(x) &= \frac{1}{2\pi(\ell\cdot x)} \ .
\end{align}

\subsection{Fixing the normalization of the Fronsdal/twistor dictionary} \label{sec:twistors:normalization}

Let's now work out the 2-point function for the spin-$s$ currents \eqref{eq:contracted_j}. By $SO(1,4)$ symmetry, it must assume the same form \eqref{eq:S_local_0}-\eqref{eq:S_local_s} as the quadratic bulk action from two boundary-to-bulk propagators. Our interest is in the normalization coefficient.

It is convenient to start from the bilocal scalar operators $\calO(\ell,\ell') = \bar\phi_I(\ell')\phi^I(\ell)$. The connected 2-point function for these is given by the Feynman diagram in figure \ref{fig:4pt}b:
\begin{align}
 \left<\calO(\ell_1,\ell'_1)\calO(\ell_2,\ell'_2)\right>_{\text{connected}} = NG(\ell'_1,\ell_2)G(\ell'_2,\ell_1) = \frac{N}{32\pi^2\sqrt{(\ell'_1\cdot\ell_2)(\ell_1\cdot\ell'_2)}} \ . \label{eq:bilocal_2_point}
\end{align}
Here, $G = \Box^{-1}$ is the propagator of the vector model's fundamental field $\phi^I$:
\begin{align}
 G(\ell,\ell') = -\frac{1}{4\pi\sqrt{-2\ell\cdot\ell'}} \ , \label{eq:G}
\end{align}
which is just the embedding-space expression for the massless propagator $G = -1/(4\pi r)$ in 3d flat space.

To obtain the 2-point function of the spin-0 local ``current'' $j^{(0)}(\ell) = \bar\phi_I(\ell)\phi^I(\ell)$, we simply set $\ell_1=\ell'_1$ and $\ell_2 = \ell'_2$ in \eqref{eq:bilocal_2_point}:
\begin{align}
 \left<j^{(0)}(\ell_1)\,j^{(0)}(\ell_2)\right>_{\text{connected}} = -\frac{N}{32\pi^2(\ell_1\cdot\ell_2)} \ . \label{eq:2_point_0}
\end{align}
For the 2-point function of currents with nonzero spin, we must act on \eqref{eq:bilocal_2_point} with derivatives according to the pattern in \eqref{eq:contracted_j}, contract with polarization vectors $\lambda^\mu_1,\lambda^\mu_2$, and then set $\ell_1=\ell'_1$ and $\ell_2=\ell'_2$ in the end. Performing this procedure on the first of the two bilocals in \eqref{eq:bilocal_2_point}, we get:
\begin{align}
 \left<j^{(s)}(\ell_1,\lambda_1)\,\calO(\ell_2,\ell'_2)\right>_{\text{connected}} = \frac{(2s)!}{(4i)^s s!}\cdot\frac{N}{32\pi^2}
   \sum_{m=0}^s (-1)^m \binom{s}{m}\frac{(\lambda_1\cdot\ell_2)^m(\lambda_1\cdot\ell'_2)^{s-m}}{(-\ell_1\cdot\ell_2)^{m+\frac{1}{2}}(-\ell_1\cdot\ell'_2)^{s-m+\frac{1}{2}}} \ . \label{eq:j_O}
\end{align}
Doing the same to the second bilocal, we will get terms of the general form:
\begin{align}
 \frac{(\lambda_1\cdot\lambda_2)^n(\lambda_1\cdot\ell_2)^{s-n}(\lambda_2\cdot\ell_1)^{s-n}}{(\ell_1\cdot\ell_2)^{2s-n+1}} \ , \label{eq:O_j_terms}
\end{align}
with coefficients that involve some unpleasant combinatoric sums. On the other hand, we know that these terms must eventually organize into the structure from \eqref{eq:S_local_s}:
\begin{align}
 \frac{(M^1_{\mu\nu} M_2^{\mu\nu})^s}{(\ell_1\cdot\ell_2)^{2s+1}} = \frac{2^s\left((\lambda_1\cdot\lambda_2)(\ell_1\cdot\ell_2) - (\lambda_1\cdot\ell_2)(\lambda_2\cdot\ell_1)\right)^s}{(\ell_1\cdot\ell_2)^{2s+1}} \ .
\end{align}
Thus, it's enough to follow just the coefficient of e.g. the $(\lambda_1\cdot\lambda_2)^s/(\ell_1\cdot\ell_2)^{s+1}$ term. This arises from acting with the $\del/\del\ell_2^\mu$ and $\del/\del\ell'^\mu_2$ derivatives just on the numerator in \eqref{eq:j_O}. The coefficient is now easy to work out, and we get:
\begin{align}
 \left<j^{(s)}(\ell_1,\lambda_1)\,j^{(s)}(\ell_2,\lambda_2)\right>_{\text{connected}} = \frac{(-1)^{s+1}(2s)!N}{2^{s+6}\pi^2}\cdot \frac{(M^1_{\mu\nu} M_2^{\mu\nu})^s}{(\ell_1\cdot\ell_2)^{2s+1}} \ . \label{eq:2_point_s}
\end{align}
Now, recall that the bulk action is related to the boundary partition function as $S = -\ln Z$. Thus, the quadratic contribution to the bulk action from two boundary insertions, which correspond to the boundary-to-bulk propagators \eqref{eq:boundary_bulk_C}, is simply minus the 2-point function \eqref{eq:2_point_0},\eqref{eq:2_point_s}. To conform with the conventions of the previous sections, we also divide by a factor of 2, so as to count each ordering of the two boundary insertions separately. We thus arrive at the bulk action as:
\begin{align}
 \begin{split}
   s=0:\quad &S[C_1,C_2] = \frac{N}{64\pi^2(\ell_1\cdot\ell_2)} \ ; \\
   s\geq 1:\quad &S[C_1,C_2] = \frac{(-1)^s(2s)!N}{2^{s+7}\pi^2}\cdot \frac{(M^1_{\mu\nu} M_2^{\mu\nu})^s}{(\ell_1\cdot\ell_2)^{2s+1}} \ .
 \end{split} \label{eq:2_point_action}
\end{align}
On the other hand, for boundary-to-bulk propagators expressed as Fronsdal fields $h_{\mu_1\dots\mu_s}$ of the form \eqref{eq:boundary_bulk} with curvature $\varphi_{\mu_1\nu_1\dots\mu_s\nu_s}$ of the form \eqref{eq:boundary_bulk_field_strength_0}-\eqref{eq:boundary_bulk_field_strength_s}, we've seen that the bulk action is given by eqs. \eqref{eq:S_local_0}-\eqref{eq:S_local_s}. Putting everything together, we arrive at the proportionality coefficients between the field strengths of canonically normalized Fronsdal fields, and those derived from $f(Y)$ via the Penrose transform:
\begin{align}
 C_{\mu_1\nu_1\dots\mu_s\nu_s}(x) = \frac{4\pi}{\sqrt{2^{s-1}N}}\,\varphi_{\mu_1\nu_1\dots\mu_s\nu_s}(x) \ . \label{eq:proportionality}
\end{align}
Eq. \eqref{eq:proportionality} holds for both zero and nonzero spins.

\section{The Didenko-Vasiliev solution and boundary bilocals} \label{sec:boundary_bilocals}

\subsection{DV particle as the bulk dual of a boundary bilocal}

In section \ref{sec:twistors:normalization}, we used the fact that the local single-trace operators \eqref{eq:contracted_j} of the boundary CFT can all be treated as singular limits of the simple \emph{bilocal} operator $\calO(\ell,\ell') = \bar\phi_I(\ell')\phi^I(\ell)$. This is the essence of the Flato-Fronsdal theorem \cite{Flato:1978qz}, which has been highlighted and exploited e.g. in \cite{Das:2003vw,Douglas:2010rc}. Now, in \cite{Neiman:2017mel}, we identified the twistor function that corresponds to the bilocal $\calO(\ell,\ell')$, in the same sense that the twistor functions \eqref{eq:kappa_s} correspond to the local currents \eqref{eq:contracted_j}. In other words, we found a linear map between bilocal boundary sources and twistor functions, such that the correlators $\left<\calO(\ell_1,\ell'_1)\dots\calO(\ell_n,\ell'_n)\right>$ are generated by the HS-algebraic functional \eqref{eq:Z}. The specific twistor function that corresponds to $\calO(\ell,\ell')$ reads:
\begin{align}
  K(\ell,\ell';Y) = \frac{1}{\pi\sqrt{-2\ell\cdot\ell'}}\exp\frac{iY\ell\ell' Y}{2\ell\cdot\ell'} \ . \label{eq:K}
\end{align}
To understand the origin of this function, we can write it as a star product of two local pieces:
\begin{align}
K(\ell,\ell';Y) = \frac{\kappa^{(0)}(\ell;Y)\star\kappa^{(0)}(\ell';Y)}{G(\ell,\ell')} = \frac{\sqrt{-2\ell\cdot\ell'}}{4\pi}\,\delta_\ell(Y)\star\delta_{\ell'}(Y) \ , \label{eq:K_kappa}
\end{align}
where $\kappa^{(0)}(\ell;Y)$ is the twistor function \eqref{eq:kappa_0} describing a local insertion of the spin-0 operator $j^{(0)}(\ell) = \bar\phi_I(\ell)\phi^I(\ell)$, and $G(\ell,\ell')$ is the fundamental propagator \eqref{eq:G}. The logic behind eq. \eqref{eq:K_kappa} is as follows. The Feynman diagrams of the free vector model's correlators are simply single loops, in which the operator insertions are connected by propagators $G(\ell,\ell')$ (see e.g. figures \ref{fig:3pt}b and \ref{fig:4pt}b). An $\calO(\ell,\ell')$ insertion in such a Feynman diagram behaves exactly like a pair of insertions $j^{(0)}(\ell),j^{(0)}(\ell')$ in sequence, but without the propagator between $\ell$ and $\ell'$. Eq. \eqref{eq:K_kappa} encapsulates this diagrammatic relationship in terms of HS algebra.

Now, what is the bulk master field that corresponds to the twistor function \eqref{eq:K_kappa}? This was also calculated in \cite{Neiman:2017mel}, as:
\begin{align}
 C(x;Y) = \frac{\pm 1}{\pi\sqrt{2[\ell\cdot\ell' + 2(\ell\cdot x)(\ell'\cdot x)]}}\exp\frac{iY[\ell\ell' + 2(\ell'\cdot x)\ell x] Y}{2[\ell\cdot\ell' + 2(\ell\cdot x)(\ell'\cdot x)]} \ . \label{eq:C_bilocal}
\end{align}
The overall sign is ambiguous, and will not play an important role. For later convenience, we will set it to $-1$ (see e.g. \cite{Iazeolla:2011cb,Iazeolla:2017vng} for a discussion of similar sign ambiguities). We now wish to make the key observation that the bulk fields contained in \eqref{eq:C_bilocal} are precisely the fields of an HS-charged particle, moving along the bulk geodesic between the boundary points $\ell'$ and $\ell$, carrying the Didenko-Vasiliev pattern of charges \eqref{eq:charge_pattern}. First, we observe that the denominators in \eqref{eq:C_bilocal} are proportional to the ``distance'' $R$ from the geodesic, as defined in \eqref{eq:R}:
\begin{align}
 C(x;Y) = -\frac{1}{\pi\sqrt{-2\ell\cdot\ell'}\,R}\exp\frac{-iY[\ell\ell' + 2(\ell'\cdot x)\ell x] Y}{2(\ell\cdot\ell')R^2} \ . \label{eq:C_bilocal_raw}
\end{align}
Let's now expand out the compact index-free notation in \eqref{eq:C_bilocal_raw}, and highlight the relevant index symmetries:
\begin{align}
 C(x;Y) = -\frac{1}{\pi\sqrt{-2\ell\cdot\ell'}\,R}\exp\frac{iY^a Y^b\gamma^{\mu\nu}_{ab}\left[\ell_{[\mu}\ell'_{\nu]} + 2(\ell'\cdot x)\ell_{[\mu} x_{\nu]} \right]}{2(\ell\cdot\ell')R^2} \ .  \label{eq:C_bilocal_interim}
\end{align}
Now, recall from \eqref{eq:Penrose_s}-\eqref{eq:Penrose_0} that the field strengths at $x$ (as opposed to their derivatives) are contained in the master field's dependence on purely chiral spinors at $x$, namely $Y = P_{\pm}(x)Y = y_{(\pm x)}$. When we make this substitution in \eqref{eq:C_bilocal_interim}, the bivector in square brackets gets projected onto the space of right-handed or left-handed bivectors at $x$. Since both of these spaces are orthogonal to $x^\mu$, the second term in the square brackets can be simply ignored. As for the first term, we recall from eq. \eqref{eq:bivector_projection} that its projection onto the space of bivectors at $x$ is proportional to the bivector $S^\perp_{\mu\nu}$ in the $tr$ plane, defined in \eqref{eq:curved_bivector}. Decomposing this into its right-handed and left-handed parts, we arrive at:
\begin{align}
 C(x;y_{(\pm x)}) = -\frac{1}{\pi\sqrt{-2\ell\cdot\ell'}\,R}\exp\frac{iY^a Y^b\gamma^{\mu\nu}_{ab}S^{R/L}_{\mu\nu}}{2R^2} \ .
\end{align}
From here, we extract the field strengths of different spins using \eqref{eq:Penrose_s}-\eqref{eq:Penrose_0}:
\begin{align}
 \begin{split}
   s=0:\quad &C(x) = -\frac{1}{\pi\sqrt{-2\ell\cdot\ell'}}\cdot\frac{1}{R} \ ; \\
   s\geq 1:\quad &C_{\mu_1\nu_1\dots\mu_s\nu_s}(x) = -\frac{i^s(2s)!}{\pi s!\sqrt{-2\ell\cdot\ell'}}\cdot\frac{S^L_{\mu_1\nu_1}\dots S^L_{\mu_s\nu_s} + S^R_{\mu_1\nu_1}\dots S^R_{\mu_s\nu_s}}{R^{2s+1}} \ .
 \end{split}
\end{align}
Finally, we use eq. \eqref{eq:proportionality} to convert these into the field strengths of canonically normalized Fronsdal fields:
\begin{align}
 \begin{split}
   s=0:\quad &\varphi(x) = -\frac{\sqrt{N}}{8\pi^2\sqrt{-\ell\cdot\ell'}}\cdot\frac{1}{R} \ ; \\
   s\geq 1:\quad &\varphi_{\mu_1\nu_1\dots\mu_s\nu_s}(x) = -\frac{i^s(2s)!\sqrt{2^s N}}{8\pi^2 s!\sqrt{-\ell\cdot\ell'}}\cdot\frac{S^L_{\mu_1\nu_1}\dots S^L_{\mu_s\nu_s} + S^R_{\mu_1\nu_1}\dots S^R_{\mu_s\nu_s}}{R^{2s+1}} \ .
 \end{split} \label{eq:bilocal_varphi}
\end{align}
We now observe that these are just the field strengths \eqref{eq:curved_solution_0},\eqref{eq:curved_C} of an HS-charged particle from section \ref{sec:bulk_particle:EAdS}, with charges:
\begin{align}
 Q^{(s)} = \frac{i^s\sqrt{2^s N}}{2\pi\sqrt{-\ell\cdot\ell'}} \ . \label{eq:charges}
\end{align}
This pattern of charges precisely agrees with the one we identified in eq. \eqref{eq:charge_pattern} as cancelling UV divergences in the two-particle interaction. 

Note that a curious thing has happened here. Normally, the Penrose transform should produce solutions to \emph{free} massless field equations, without bulk sources. For the boundary-to-bulk propagators \eqref{eq:kappa_s},\eqref{eq:boundary_bulk_C} corresponding to the local boundary operator $j^{(s)}(\ell,\lambda)$, this is indeed the case. However, we now see that the Penrose transform of the twistor function \eqref{eq:K} solves not quite the free linearized equations, but the equations with a \emph{particle-like source} along the bulk geodesic between $\ell'$ and $\ell$. This puts us into somewhat new territory for holography. In particular, one may wonder: are the boundary 2-point correlators still described by a quadratic bulk action, even though the corresponding bulk fields are no longer free?

It turns out that the answer is yes, provided we define the bulk action as in \eqref{eq:S_with_particle}, including both the free-field term and the interaction term with the bulk ``particle''. Indeed, consider two boundary bilocals, $\calO(\ell_1,\ell'_1)$ and $\calO(\ell_2,\ell'_2)$. Each of these generates bulk HS fields, which are the fields of an HS-charged particle with charges given by \eqref{eq:charges}. We can then use the result \eqref{eq:curved_result} of section \ref{sec:two_particles} to evaluate the quadratic bulk action as:
\begin{align}
 S = -\frac{N}{64\pi^2\sqrt{(\ell_1\cdot\ell'_2)(\ell_2\cdot\ell'_1)}} \ . \label{eq:bilocal_2_point_again}
\end{align}
This is $-\frac{1}{2}$ times the correlator \eqref{eq:bilocal_2_point} of the two bilocals, in agreement with the holographic dictionary (recall eqs. \eqref{eq:2_point_0},\eqref{eq:2_point_s} as compared to eq. \eqref{eq:2_point_action}). Note that if we were to consider only the first, ``free-field'' term in the action \eqref{eq:S_with_particle}, the result would have the opposite sign. Thus, the interaction term with the bulk ``particle'' must be included \emph{both} in the field equations for the linearized HS fields (otherwise \eqref{eq:bilocal_varphi} is not a solution), \emph{and} when evaluating the bulk action (which otherwise fails to agree with the bilocal correlator \eqref{eq:bilocal_2_point}). In other words, if we wish to work with boundary bilocals, we have no choice but to account for the existence of DV particle-like sources in the bulk.

Note that the action \eqref{eq:bilocal_2_point_again} is divergent only when the boundary endpoints $(\ell_1,\ell'_2)$ or $(\ell_1',\ell_2)$ coincide. When the geodesics of the DV particles intersect \emph{in the bulk}, the action is perfectly regular; this is the cancellation of bulk UV divergences discussed in section \ref{sec:two_particles}. In particular, the action for a DV particle interacting \emph{with itself}, such that $\ell_1=\ell_2$ and $\ell_1'=\ell_2'$, is finite. On the other hand, if we restrict to even spins by symmetrizing over $(\ell_1\leftrightarrow\ell_1')$ and $(\ell_2\leftrightarrow\ell_2')$, this self-interaction becomes divergent.

\subsection{Relation to the Didenko-Vasiliev ``black hole''} \label{sec:boundary_bilocals:comparing_formalisms}

So far, we've shown that the linearized HS fields of a ``Didenko-Vasiliev particle'', as defined in section \ref{sec:two_particles}, coincide with the bulk fields that correspond to a bilocal operator in the boundary CFT. In this section, we observe that they \emph{also} coincide with the linearized version of the Didenko-Vasiliev black hole, thus justifying our nomenculature. Our statement is that the bulk master field \eqref{eq:C_bilocal}, derived via the Penrose transform \eqref{eq:Penrose} from the twistor function \eqref{eq:K}, is the same as the linearized Didenko-Vasiliev solution as given in \cite{Didenko:2009td}, up to slight differences in the formalism (and in the spacetime signature). 

First, let us summarize the differences and similarities between the twistor formalism presented here and the one found in ``mainstream'' HS literature. The formalism in this paper, which was first put forward in \cite{Neiman:2015wma}, starts with a fixed $EAdS_4$ geometry, defined via an $\bbR^{1,4}$ embedding space. The tangent space at a spacetime point $x\in EAdS_4$ is just the tangent 4d hyperplane in $\bbR^{1,4}$ to the $EAdS_4$ hyperboloid; the tangent spaces for different points are represented by different 4d hyperplanes in the same $\bbR^{1,4}$. Twistor space is defined as the space of $SO(1,4)$ spinors. At a point $x\in EAdS_4$, it decomposes into two spaces of Weyl $SO(4)$ spinors; the Weyl spinor spaces at different points are represented by different 2d subspaces of the same twistor space. 

In contrast, in the standard HS literature, one doesn't have a fixed $EAdS_4$ geometry or an embedding space, but an a-priori featureless spacetime manifold. On it, one constructs the frame fields of Cartan's formulation of General Relativity, and their higher-spin extensions. Thus, the tangent space and Weyl spinor spaces at different spacetime points $x$ exist only as fibers over the spacetime manifold, as is usually the case in GR. The left-handed and right-handed spinors at $x$ are unified into Dirac spinors $Y$. These are referred to as ``twistors'', but ``only'' due to the structure imposed on them by HS algebra, acting on the fiber at $x$. There is no notion of a twistor $Y$ that exists independently from the spacetime point $x$. Again, this is as usual in GR: true, Penrosian, $x$-independent twistors are easy to define only on very special spacetimes.

With these basic circumstances in mind, let us consider again the HS-algebraic Penrose transform \eqref{eq:Penrose}. The bulk master field $C(x;Y)$ on the LHS of \eqref{eq:Penrose} is basically the same as that in the standard HS literature, up to the aforementioned difference in the nature of Weyl spinors at $x$: in the standard formalism, they are basic structures in the fiber at $x$, while in ours, they are $x$-dependent projections of an $x$-independent twistor $Y$. The same comments apply to the spinor delta function $\delta_x(Y)$ on the RHS of \eqref{eq:Penrose}. As for the $x$-independent twistor function $f(Y)$ on the RHS of \eqref{eq:Penrose}, one may think at first that it has no analog in the standard HS formalism. And yet, essentially the same formula as \eqref{eq:Penrose} was put forward in eq. (3.23) of \cite{Didenko:2009td}, as a technique for generating free bulk solutions. Instead of a twistor function $f(Y)$ that's literally \emph{constant} with respect to $x$, in \cite{Didenko:2009td} one uses a function $\epsilon_0(x;Y)$ that is \emph{covariantly constant} with respect to the HS connection, in the \emph{adjoint representation} of HS symmetry. A star product with a spinor delta function, just as in \eqref{eq:Penrose}, transforms this function into a master field that solves the linearized bulk equations, and in turn lives in the so-called \emph{twisted adjoint} representation of HS symmetry. Upon some reflection, one can see that eq. (3.23) of \cite{Didenko:2009td} and our Penrose transform \eqref{eq:Penrose} are really the same, up to the above ``cosmetic'' differences in formalism. In particular, as was shown in \cite{Neiman:2017mel}, our twistor function $f(Y)$ lives in the adjoint representation of HS symmetry just like the $\epsilon_0(x;Y)$ of \cite{Didenko:2009td}, while our master field $C(x;Y)$ lives in the twisted adjoint.

Now, the authors of \cite{Didenko:2009td} proceeded to construct a \emph{particular} solution to the linearized bulk equations -- the linearized Didenko-Vasiliev black hole -- out of a particular covariantly constant twistor function $\epsilon_0(x;Y)$:
\begin{align}
 \epsilon_0(x;Y) \sim e^{i\calK_{ab}(x)Y^a Y^b/2} \ . \label{eq:epsilon}
\end{align}
Here, $\calK_{ab}(x)$ a generator of the $AdS_4$ group -- specifically, the generator of time translations in the black hole's rest frame -- normalized as:
\begin{align}
 \calK^a{}_b \calK^b{}_c = \delta^a_c \ . \label{eq:generator_normalization}
\end{align}
In the linearized limit, when we consider the ``black hole'' as a point particle, this is just the generator of time translations along the particle's geodesic worldline. Now, consider the embedding space $\bbR^{2,3}$ of (now, Lorentzian) $AdS_4$. There, the particle's worldline is just the intersection of the $AdS_4$ hyperboloid with a 2d plane through the origin of $\bbR^{2,3}$, spanned by some simple bivector $S_{\mu\nu}$. We then recognize $\calK_{ab}\sim \gamma_{ab}^{\mu\nu}S_{\mu\nu}$ as the generator of rotations in this 2d plane. In the embedding-space formalism, this generator is $x$-independent. 

All that remains now is to switch signatures to $EAdS_4$, with $\bbR^{1,4}$ embedding space. The particle's worldline becomes a \emph{spacelike} geodesic, with boundary endpoints $\ell$ and $\ell'$, such that $S_{\mu\nu}\sim \ell_{[\mu}\ell'_{\nu]}$. Imposing the normalization condition \eqref{eq:generator_normalization}, we get:
\begin{align}
 \calK_{ab} = \pm \frac{\gamma_{ab}^{\mu\nu}\ell_\mu\ell'_\nu}{\ell\cdot\ell'} \ .
\end{align}
We thus see that, upon translation to the present paper's formalism, the twistor function \eqref{eq:epsilon} from \cite{Didenko:2009td} is nothing but our twistor function \eqref{eq:K} that corresponds to the boundary bilocal! Therefore, its Penrose transform \eqref{eq:C_bilocal}, which solves the linearized field equations with a particle-like source with charges \eqref{eq:charges}, is just the linearized version of the Didenko-Vasiliev black hole from \cite{Didenko:2009td}. This justifies our terminology of referring to particles with the pattern of charges \eqref{eq:charge_pattern} as ``DV particles''.

\section{Discussion} \label{sec:discuss}

In this paper, we pointed out two new perspectives on the Didenko-Vasiliev ``black hole'' solution, or rather its linearized version. First, we learned to view this solution in terms of Fronsdal fields generated by an HS-charged particle, with the special pattern of charges \eqref{eq:charge_pattern}. We calculated the interaction between such two particles via their HS fields, and found that the DV pattern of charges has a unique property: in a certain sense, it makes the two-particle interaction non-local. Specifically, with this pattern of charges, the interaction action does not have a short-distance singularity as the two particles are brought close together, for almost any angle between the worldlines; the exception is the angle $\theta = \pi$, which in Lorentzian corresponds to a particle and antiparticle mutually at rest. 

Second, we learned to identify the DV solution as the bulk holographic dual of a boundary bilocal operator $\calO(\ell,\ell') = \bar\phi_I(\ell')\phi^I(\ell)$, and showed that for two such objects, the bulk interaction action agrees with the connected boundary correlator. In more detail, we saw that a boundary bilocal operator generates a bulk DV particle that travels along the geodesic between the two boundary points, and carries HS charges that source the bulk HS gauge fields; the correlator of two boundary bilocals can then be expressed as the exchange of HS fields between the two bulk DV particles. Though such a picture is new to HS theory, it's actually been painted before within the general AdS/CFT (or even general CFT) context. We are referring here to \emph{geodesic Witten diagrams}, introduced in \cite{Hijano:2015zsa,daCunha:2016crm} as a bulk representation for boundary OPE blocks and conformal blocks. Indeed, one way of phrasing the Flato-Fronsdal theorem is that the bulk HS multiplet (or the boundary multiplet of HS currents) is nothing but the OPE of the fundamental boundary fields $\phi^I(\ell)$ and $\bar\phi_I(\ell)$ (with the identity operator excluded). Thus, our picture of a bulk geodesic stretching between $\ell$ and $\ell'$ and sourcing the HS multiplet is precisely the ``half-geodesic Witten diagram'' picture of OPE blocks proposed in \cite{daCunha:2016crm} (see figure \ref{fig:3pt}). Similarly, the connected correlator \eqref{eq:bilocal_2_point} of two bilocals can be thought of as a contribution to the 4-point function of the fundamental boundary fields; our picture of it as an HS field exchange between two geodesics (see figure \ref{fig:4pt}) is precisely the geodesic Witten diagram description of conformal blocks proposed in \cite{Hijano:2015zsa}.

That being said, we view our results as more than just a special case of the geodesic Witten diagram framework. This is because of their particularly tight relation to the basic structure of HS theory. The bulk DV particle embodies not just \emph{some} OPE, but \emph{the} OPE that defines the entire spectrum of HS theory. Furthermore, the role of the fundamental fields $\phi^I,\bar\phi_I$ here is subtle: they carry $U(N)$ color, and therefore aren't usually considered as part of the CFT's operators. As a result, the DV particle that embodies their bulk OPE is really a new ingredient in the bulk theory. It is telling us that, if we wish to accommodate boundary bilocals, then HS gravity must include more than just HS gauge fields interacting with each other: we must also allow for particle-like HS currents that act as \emph{bulk sources} for the HS fields. 

To make this more concrete, let us point out a specific property that sets the DV particle apart from the HS gauge multiplet. Unlike any of the HS gauge fields, the DV particle carries \emph{electric charge}: it is charged under the spin-1 gauge field in the HS multiplet, or, equivalently, under the $U(1)$ part of the boundary color $U(N)$. This is easy to understand from the boundary perspective: there, the electric charge is carried by $\phi^I$, while $\bar\phi_I$ carries an opposite charge. The local HS currents \eqref{eq:j}, which are the holographic duals of the usual bulk fields, are all electrically neutral, since they contain a product of $\phi^I(\ell)$ and $\bar\phi_I(\ell)$. In contrast, the bilocal operator $\calO(\ell,\ell') = \bar\phi_I(\ell')\phi^I(\ell)$ is positively charged at one point, and negatively charged at another; in the bulk, this translates into an electrically charged DV particle that travels between the two points. 

Now, suppose that we take seriously the possibility of boundary bilocal operators, and with them their corresponding bulk HS currents. Then we should ask: what is the general form of such currents? We've seen that a single bilocal insertion produces a particle-like current \eqref{eq:T} with the DV pattern of charges \eqref{eq:charge_pattern}. What about a general superposition of such insertions? Does the DV pattern of charges for each insertion entail some restrictions on the resulting bulk currents? It's easy to see that it does. The relations \eqref{eq:charge_pattern} between the charges of different spins translate into a linear relation between the corresponding currents, which will be preserved by superpositions. Specifically, the trace of the spin-$(s+2)$ current $T^{\mu_1\dots\mu_{s+2}}$ ends up proportional to the traceless part of the spin-$s$ current $T^{\mu_1\dots\mu_s}$:
\begin{align}
 T_\nu^{\nu\mu_1\dots\mu_s} = -2\left(T^{\mu_1\dots\mu_s} - \frac{s-1}{4}\,g^{(\mu_1\mu_2}T_\nu^{\mu_3\dots\mu_s)\nu} \right) \ . \label{eq:currents_pattern}
\end{align}
This relation generalizes the DV pattern of charges \eqref{eq:charge_pattern} to continuous current distributions in the bulk. It implies that, while the currents at each point $x$ are merely \emph{double-}traceless, only their fully traceless parts are independent. We conjecture that any configuration of bulk HS currents that satisfies eq. \eqref{eq:currents_pattern} (and the appropriate conservation laws) arises from some superposition of boundary bilocals. For the bulk HS fields, the constraint \eqref{eq:currents_pattern} on the currents translates into a relation between the Fronsdal tensors of different spins:
\begin{align}
 F_\nu^{\nu\mu_1\dots\mu_s} = \frac{2}{s+1}\left(F^{\mu_1\dots\mu_s} - \frac{s-1}{4}\,g^{(\mu_1\mu_2}F_\nu^{\mu_3\dots\mu_s)\nu} \right) \ . \label{eq:F_pattern}
\end{align}
Thus, if we allow arbitrary bilocal sources, the spectrum of bulk HS fields gets effectively extended, by relaxing the linearized field equations from $F^{\mu_1\dots\mu_s} = 0$ to \eqref{eq:F_pattern}. Note that it's not clear how to express a general solution of \eqref{eq:F_pattern} as either a twistor function $f(Y)$ or a master field $C(x;Y)$. Indeed, the twistor/spinor language was sufficiently flexible to allow for a \emph{single} DV particle-like source, for which the free HS field equations are satisfied \emph{almost} everywhere (see \cite{Iazeolla:2017vng,DeFilippi:2019jqq,Aros:2019pgj} for more detailed discussions on how the twistor/spinor language bypasses or resolves the $r=0$ singularity). However, it does not seem flexible enough to accommodate a general superposition of such sources, for which the free equations aren't satisfied \emph{anywhere}. This makes for an apparent failure of linearity, similar to the one we analyzed in \cite{David:2020ptn}, and may benefit from closer attention.

Overall, what we find exciting is that, even though we dealt here only with \emph{linearized} HS gravity, the above discussion rhymes with some central issues in the interacting theory:
\begin{enumerate}
 \item In HS theory, which is described via equations of motion rather than an action, it is natural to express the interactions as a coupling between the HS fields and some effective HS currents, which are in turn non-linear combinations constructed from the HS fields. 
 \item Beginning from the quartic vertex, the interactions of HS gravity suffer from a non-locality problem \cite{Sleight:2017pcz}. In particular, the boundary scalar 4-point function $\left<j^{(0)}(\ell_1)j^{(0)}(\ell_2)j^{(0)}(\ell_3)j^{(0)}(\ell_4)\right>$ implies a non-local bulk vertex. When this problem was first glimpsed in \cite{Fotopoulos:2010ay}, it was suggested that the solution may be to include additional degrees of freedom in the description of the theory. 
 \item An attempt to address the locality issue is underway \cite{Gelfond:2018vmi,Didenko:2018fgx,Didenko:2019xzz,Gelfond:2019tac}, with so-called spin locality replacing ordinary spacetime locality as a guiding principle. As pointed out in \cite{Gelfond:2019tac}, this new locality principle is actually ordinary spacetime locality, but with the set of field variables extended to include all possible non-linear currents (but not their derivatives). Thus, the spin-locality effort is in some sense a concrete realization of the vague proposal in \cite{Fotopoulos:2010ay} to restore locality by extending the set of degrees of freedom.
\end{enumerate}
Now, in the present paper, we also extended the spectrum of bulk HS fields by including bulk HS currents. We then expressed the 2-point correlator $\left<\calO(\ell_1,\ell'_1)\calO(\ell_2,\ell'_2)\right>$ of boundary bilocals as a \emph{local bulk process} involving both these currents and the original HS fields. Most tantalizingly, the boundary Feynman diagram for this correlator is very similar to those of the infamous 4-point correlator $\left<j^{(0)}(\ell_1)j^{(0)}(\ell_2)j^{(0)}(\ell_3)j^{(0)}(\ell_4)\right>$, just with two of the propagators removed (see eqs. \eqref{eq:bilocal_2_point},\eqref{eq:K_kappa} and figure \ref{fig:4pt}). This leads us to hope that the construction presented here may provide a useful alternative viewpoint on the bulk interactions and their locality properties.

\section*{Acknowledgements}

We are grateful to Eugene Skvortsov, Per Sundell and Mirian Tsulaia for discussions. We are especially grateful to Slava Lysov for helping us realize the full extent of the equivalence between the DV charges and the cancellation of divergences. This work was supported by the Quantum Gravity and Mathematical \& Theoretical Physics Units of the Okinawa Institute of Science and Technology Graduate University (OIST). YN's thinking was substantially informed by talks and discussions at the workshop ``Higher spin gravity -- chaotic, conformal and algebraic aspects'' at APCTP in Pohang.

\end{document}